\documentclass[sn-mathphys,iicol]{sn-jnl}

\usepackage{placeins}
\usepackage{caption}
\usepackage{subcaption}
\usepackage{ragged2e}
\usepackage{comment}

\jyear{2023}%

\theoremstyle{thmstyleone}%
\theoremstyle{thmstyletwo}%

\theoremstyle{thmstylethree}%

\begin{document}

\title[Unsteadiness]{Study of Unsteadiness due to 3-D Shock-Boundary Layer Interaction in Flow over a Square-faced Protuberance}


\author*[1]{\fnm{Ramachandra} \sur{K}}\email{ae20s003@smail.iitm.ac.in}

\author[2]{\fnm{Sourabh} \sur{Bhardwaj}}\email{sourabhb@kth.se}

\author[3]{\fnm{Jayaprakash} \sur{N. Murugan}}\email{jayaprakash.n@vit.ac.in}

\author[4]{\fnm{Sriram} \sur{R}}\email{r.sriram@iitm.ac.in}

\affil*[1]{\orgdiv{Department of Aerospace Engineering}, \orgname{Indian Institute of Technology Madras}, \orgaddress{\city{Chennai}, \postcode{600 036}, \state{Tamil Nadu}, \country{India}}}

\affil[2]{\orgdiv{Department of Engineering Mechanics}, \orgname{KTH Royal Institute of Technology}, \orgaddress{\city{Stockholm}, \postcode{114 28}, \country{Sweden}}}

\affil[3]{\orgdiv{School of Mechanical Engineering}, \orgname{Vellore Institute of Technology}, \orgaddress{\city{Vellore}, \postcode{\\ 632 014}, \state{Tamil Nadu}, \country{India}}}

\affil[4]{\orgdiv{Department of Aerospace Engineering}, \orgname{Indian Institute of Technology Madras}, \orgaddress{\city{Chennai}, \postcode{600 036}, \state{Tamil Nadu}, \country{India}}}


\abstract{The dynamics of shock-induced unsteady separated flow past a three-dimensional square-faced protuberance is investigated through wind tunnel experiments. Time-resolved schlieren imaging and unsteady surface pressure measurements are the diagnostics employed. Dynamic Mode Decomposition (DMD) of schlieren snapshots, and analysis of spectrum and correlations in pressure data are used to characterize and resolve the flow physics. The mean shock foot in the centreline is found to exhibit a Strouhal number of around 0.01, which is also the order of magnitude of the Strouhal numbers reported in the literature for two-dimensional shock-boundary layer interactions. The wall pressure spectra, in general, shift towards lower frequencies as we move away from (spanwise) centreline with some variation in the nature of peaks. The cross-correlation analysis depicts the strong dependence of the mean shock oscillations to the plateau region, and disturbances are found to travel upstream from inside the separation bubble. Good coherence is observed between the spanwise mean shock foot locations till a strouhal number of about 0.015 indicating that the 3-D shock foot largely moves to-and-fro in a coherent fashion.}




\maketitle

\section{Introduction}\label{sec1}

Wall-bounded high-speed flows often encounter Shock-Boundary Layer Interactions (SBLI). SBLI can occur in multitude of scenario such as normal/oblique shock impinging on boundary layer, interaction of compression corner/protuberance shock with incoming boundary layer, or shock occurring over walls due to imposed pressure conditions. SBLIs are associated with thickening of the boundary layer and possible flow separation depending upon the strength of the shock interacting with the boundary layer. The separated flow field is inherently unsteady, resulting in fluctuating pressure loads and peak heat transfer rates, causing detrimental effects to the systems in which they occur \cite{Babinsky, Delery}.

Research in this realm has been carried out for more than 6 decades. Earliest works concerned quantifying mean flow variables in the interaction regime, thus developing an understanding of the different zones and features that constitute the interaction region. A number of canonical configurations of interest (both 2-D and 3-D) such as compression corners \cite{Burggraf}, unswept impinging shock \cite{Green}, swept impinging shock \cite{Settles}, cylinders \cite{Ozcan, Dolling}, blunt/sharp fins \cite{Hung}, step in the flow \cite{Zukoski} etc., were explored in detail. With regard to 3-D interactions due to surface protuberances, Ozcan and Holt \cite{Ozcan} carried out experimental investigations of laminar interaction of supersonic flow over cylinders with varying H/D ratios, discussed the surface streakline pattern and compared the separation extents. The disagreement of the velocity measurements with that of the mean flow structure strongly suggested the presence of unsteadiness which was also subsequently reported in the computations performed by Lakshmanan and Tiwari \cite{Lakshmanan}. The latter work also confirmed the presence of a number of vortices present in the separation regime. Many other older works that extensively studied the separation scales such as \cite{Sedney, Hung, Dolling} also stressed on the significance of quantifying the dynamics associated with the shock foot.

The source of unsteadiness in SBLI associated with simple 2-D configurations have been widely studied by researchers. The shock oscillation in such cases is attributed to two main causes - the upstream causes from the turbulent scales in the incoming boundary layer and the downstream causes, from within the separated region itself. In the interactions with laminar boundary layers, it is only the downstream mechanism which is responsible for the observed unsteadiness. Loth and Matthys \cite{Loth} studied low Reynolds number reflected shock interactions using unsteady finite element computations and observed the separation region to be unsteady along with the occurence of eddy shedding for Reynolds numbers of over 9600 (defined based on the distance of shock impingement point from the leading edge). Robinet \cite{Robinet} studied the effect of increasing shock impingement angle in laminar separation using 3-D direct numerical simulations and demonstrated the unsteadiness through a linearized global instability analysis.

Andreopoulos and Muck \cite{Andreopoulos} presented a strong case for the upstream causes of unsteadiness in turbulent SBLI. They carried out measurements of pressure fluctuations in the interaction region of Mach 2.9 compression ramp flow and found that the shock motions are of the same frequency as that of the bursting phenomena in the upstream boundary layer. The PIV measurements by Beresh et al. \cite{Beresh} identified that the mean velocity profile was fuller during downstream shock sweep and vice-versa, signifying the role of upstream mechanism. Some of the other earlier works that supported this mechanism were by Plotkin \cite{Plotkin}, Hou \cite{Hou} etc. Later, Ganapathisubramani et al. \cite{Ganapathisubramani} gave supporting results when they identified that the turbulent superstructures in the incoming boundary layer could be the cause of the low frequency motion. Contradictorily, the experimental observations by Humble et al. \cite{Humble} and direct numerical simulations by Wu and Martín \cite{Wu} identified that the superstructures were only responsible for the spanwise wrinkling of the separation line which is characterized by small amplitudes and high frequency, while the span-averaged large amplitude streamwise oscillations are governed by the downstream flow or by the low-frequency content in the incoming flow.

On the other hand, some researchers strongly argued that the downstream mechanism is the responsible factor. The argument of shock oscillations being primarily driven by burst-sweep events in the upstream boundary layer by Andreopoulos and Muck \cite{Andreopoulos} was contradicted by the experimental observations of Thomas et al. \cite{Thomas} and they rather identified a strong relation of the shock dynamics with that of the inherent unsteadiness inside the bubble. Dupont et al. \cite{Dupont} studied the impinging shock-turbulent boundary layer interactions and observed high coherence in the low frequency regime between the reflected shock and the separated region, concluding that the unsteadiness is primarily driven by the downstream separation. Piponniau et al. \cite{Piponniau} proposed a satisfactory model based on the imbalance between the entrainment of fluid by the shear layer and subsequent recharge towards the downstream reattachment location resulting in the motion of the bubble. Similarly, multiple works \cite{Touber, Dussauge, Priebe} have added to the previous conclusions supporting the downstream mechanism. The present consensus with regard to 2-D interactions, as discussed by Souverein et al \cite{Souverein}, is that the downstream mechanism dominates for strongly separated flows whereas the incoming boundary layer scales also contribute significantly for weak interactions. Recently, Murugan and Govardhan \cite{Murugan} studied Mach 2.54 flow over a forward facing step using wall pressure measurements and PIV which identified significant correlation of the shock with both upstream and downstream regions. A detailed review of the works on 2-D configurations have been reported in \cite{Clemens}.

\begin{figure*}[hbt!]
\centering
\includegraphics[width=0.85\textwidth]{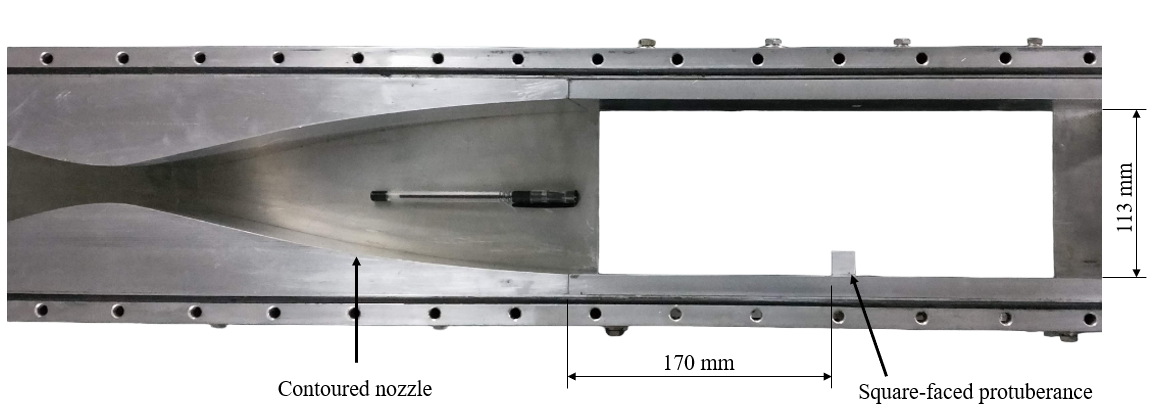}
\caption{Photograph of the test section with the nozzle}
\label{test_section}
\end{figure*}

In general, all SBLIs are three-dimensional irrespective of the source causing the shock since the flow is always associated with spanwise variations like corrugations and ripples \cite{Muck}. Here, we aim to study a scenario in which the source itself is a 3-dimensional protuberance placed on the wall, causing an inviscid bow shock - effectively having a variation in shock strength along the span - to interact with the boundary layer. Unlike nominally 2-dimensional configurations, a generalized understanding for SBLI due to 3-dimensional configurations cannot be attempted due to the myriad configurations as well as the complex nature of the flow. Earlier works have studied interactions associated with three-dimensional configurations such as sharp-fins \cite{Pickles}, blunt-fins \cite{Brusniak,Hung}, hemisphere \cite{Wang} etc. Pickles et al. \cite{Pickles} explained the spanwise relief associated with three-dimensional interactions due to sharp fins. The work by Brusniak and Dolling \cite{Brusniak} and the recent work on transitional SBLI by Murphree et al. \cite{Murphree} presented detailed correlation and spectral analyses of fluctuating pressure for blunt-fin interactions but the measurements were only along the centreline. Hung and Buning \cite{Hung} reported spanwise data of turbulent SBLI in blunt-fins but they were only reliable for mean flow variables since it involved RANS computations. In essence, understanding the overall physics of unsteadiness in 3-D configurations still remains a big challenge. This is due to the complex nature of the separated flow that is driven by several additional factors as compared to that of the two-dimensional counterparts. 

In this work, experimental analyses of flow past a square-faced obstacle protruding out of the turbulent boundary layer have been carried out. Bhardwaj et al. \cite{Sourabh} have recently carried out a detailed study of mean flow variables on multiple geometries of protuberances and have arrived at a universal scaling law that predicts the separation extent for the 2-D counterpart as an asymptotic limit. With the support of the mean flow data by \cite{Sourabh}, time-resolved schlieren imaging and unsteady surface pressure measurements have been carried out in the study with the aim of understanding the spanwise organization of the 3-D flow dynamics and provide comparative conclusions with that of the two-dimensional configuration. Modal analyses of the schlieren snapshots, spectral and correlation analyses of wall pressure data have been carried out in order to study the flow physics.

The paper is organized as follows: The experimental setup and methodology used for flow diagnostics have been discussed in section \ref{exp_setup}. The results of shock foot unsteadiness analyses using scan line and Dynamic Mode Decomposition (DMD) performed over schlieren images have been described in section \ref{shock_unsteadiness}. This is followed by a detailed discussion on the space-time flow organization using fluctuating wall pressure measurements in the section \ref{unsteady_pressure}. Conclusions drawn based on the analyses have been presented in the final section.

\section{Experimental setup and methodology} \label{exp_setup}

\begin{figure*}[hbt!]
\centering
\includegraphics[width=\textwidth]{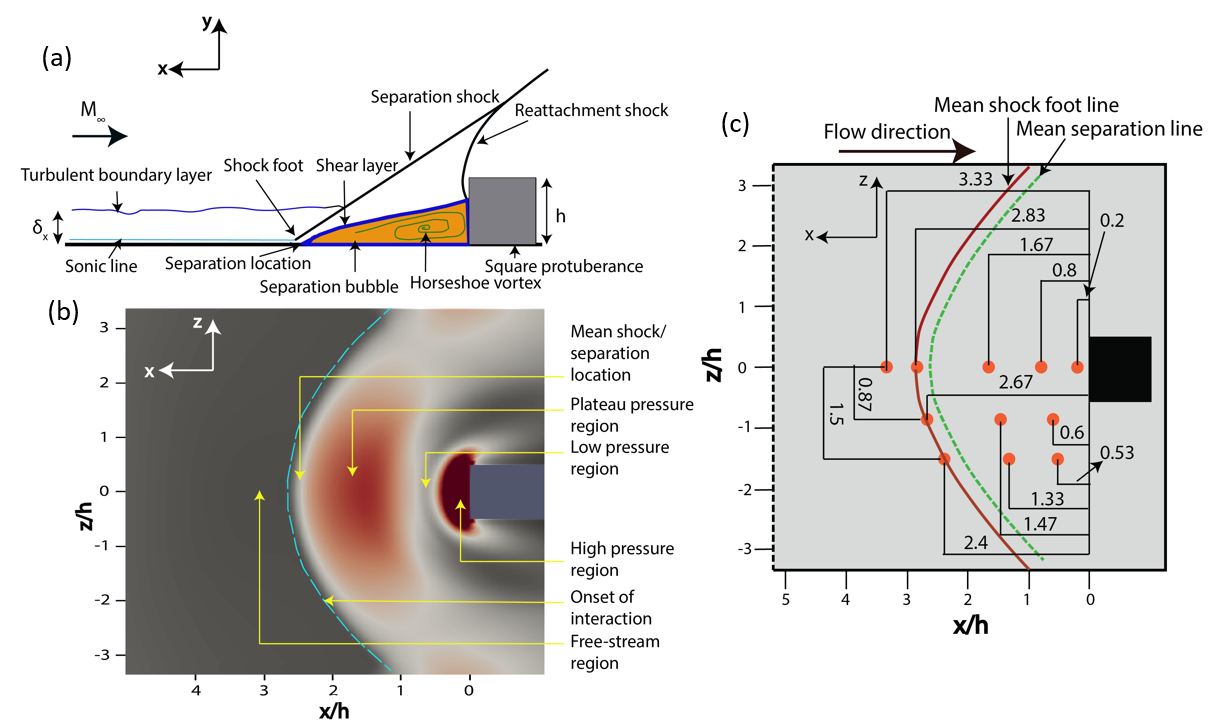}
\caption{\textbf{Left:} Schematic of the overall SBLI flow field along with different pressure zones indicated; \textbf{Right:} Locations of pressure ports used for unsteady measurements at different zones}
\label{schematic}
\end{figure*}

The experiments were conducted in the blow-down supersonic wind tunnel facility at the Gas Dynamics Laboratory, IIT Madras. The stagnation pressure and temperature were 6 bar (absolute) and 300 K respectively. The nominal Mach number in the test section was 2.87 and the boundary layer thickness ($\delta_x$) was measured to be 7 mm (using a pitot survey) at a distance of 170 mm from the nozzle exit where the front face of the protuberance was positioned. The test section has a rectangular cross section of 100 mm x 113 mm and extends for a length of 396 mm. The Reynolds number based on boundary layer thickness ($Re_{\delta}$) was found to be $2.8$ x $10^5$. The photograph of the test section along with the contoured nozzle used for the experiments is shown in fig. \ref{test_section}. More details of the setup have been discussed by Bhardwaj \cite{Bhardwaj}. Since highly blunted bodies such as flat-faced ones result in large shock stand-off distances and bow shock radii, which in turn would result in large interaction length scales that can be easily resolved with diagnostics, a square-faced protuberance was chosen for the study. The protuberance mounted in the test section is a cubical block of side 15 mm with one of its faces, oriented perpendicular to the flow direction. The height/width of the face of the protuberance is thus roughly twice the boundary layer thickness ($\delta_x$).

A schematic of the SBLI associated with the square-faced protuberance (in the spanwise centreplane) is shown in fig. \ref{schematic}(a). Figure. \ref{schematic}(b) shows another perspective of the flow field based on surface pressure distribution obtained from the RANS computations by Bhardwaj et al. \cite{Sourabh}. The protuberance causes an inviscid bow shock which interacts with the boundary layer resulting in separation. This acts as a compression corner and creates the separation shock that interacts with the reattachment bow shock causing Edney type VI interaction \cite{Edney}. 

Based on the wall pressure distribution, different zones in the bow shock induced separated flow field were identified by Bhardwaj et al. \cite{Sourabh}. Going from left to right (i.e., in streamwise direction) in fig. \ref{schematic}(b), the pressure starts to rise from the free-stream values at the location of the onset of interaction, whose locus is shown as a thin dotted line. The thickening of the boundary layer from the onset of interaction results in a series of compression waves which coalesce to form the separation shock. The pressure thus continues to rise downstream until it reaches a plateau pressure after flow separation. This plateau pressure region is followed by a local dip in pressure forming the low pressure region. Subsequently, due to flow stagnation, the wall pressure reaches high values forming the high pressure region in the vicinity of the protuberance. It can be seen that the high pressure region does not extend far from the protuberance in the spanwise direction. The low pressure region is a distinctive feature of three-dimensional protuberance configurations which is absent in 2-D interactions \cite{Delery}. This region is formed due to the presence of horseshoe vortex in front of the protuberance spiralling in the spanwise directions away from the centreline, as opposed to closed streamlines in 2-D separation bubble. The above description of the different pressure zones would be of use in subsequent discussions since the placement of the fast response sensors for unsteady pressure measurements, was based on identification of these zones. Figure. \ref{schematic}(c) shows the distances of the locations of the unsteady pressure measurements from the front face of the protuberance, non-dimensionalized with respect to its height. In fig. \ref{schematic}(c), the mean separation line as obtained from the oil flow visualization and the mean shock foot line upstream of the separation line are also marked.

\subsection{Flow diagnostics}

\begin{figure*}[hbt!]
\centering
\includegraphics[width=0.85\textwidth]{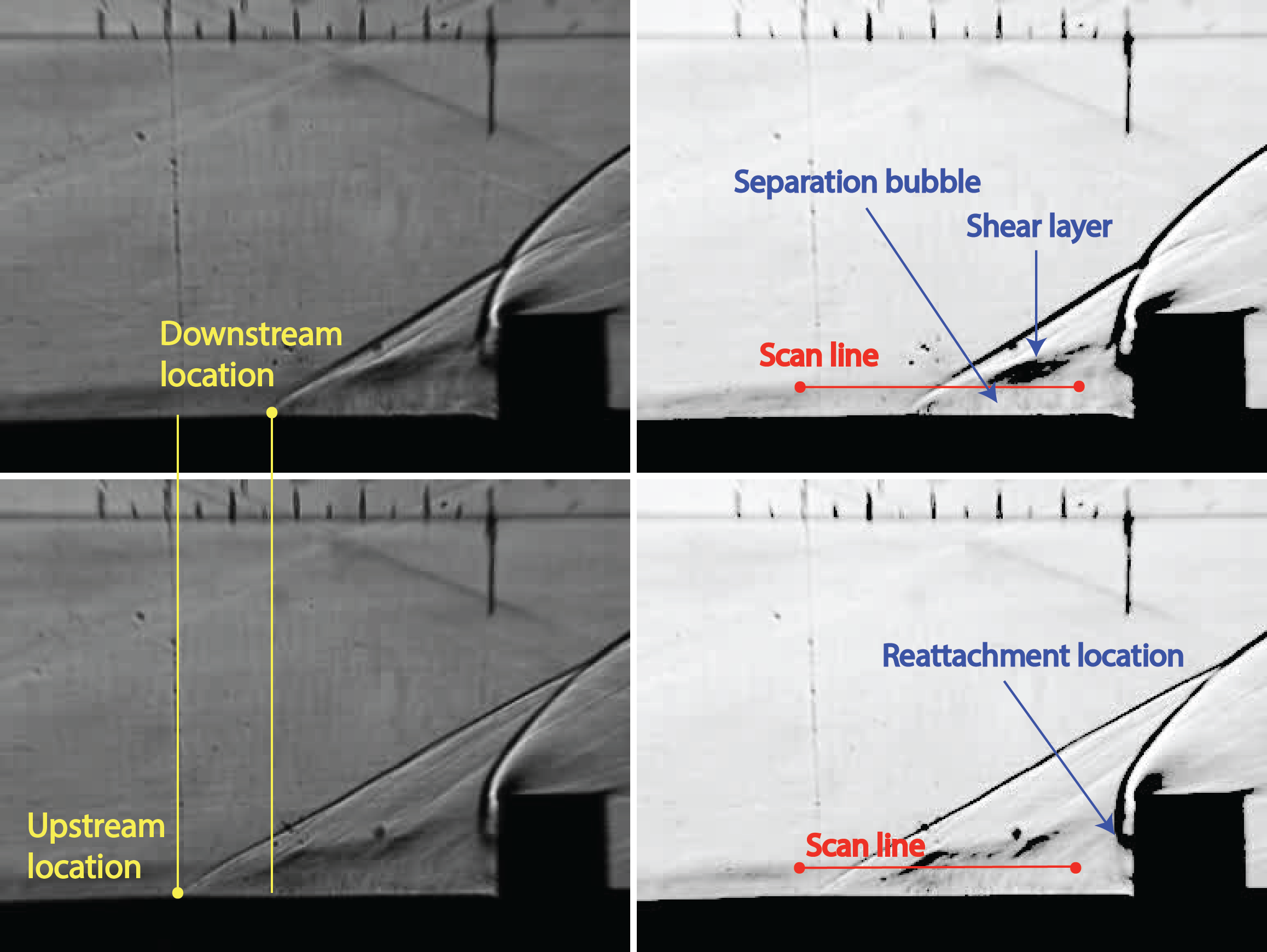}
\caption{Raw schlieren images (left) vs the processed images (right) for analysis}
\label{schlieren}
\end{figure*}
    Time-resolved schlieren imaging was performed using high-speed imaging camera (Photron FASTCAM SA4 Model 500K-M1) at a frame rate of 30000 fps with a spatial resolution and shutter speed of 384 x 288 and 1/35000 s respectively. The setup consisted of two parabolic concave mirrors of focal length 2 m (collimating mirror) and 1 m (focusing mirror), a halogen light source, a rectangular slit and a knife edge. To provide optical access, two 260 mm × 120 mm glass windows were attached to the side walls of the test section (refer \cite{Bhardwaj} for details on the setup). Fig. \ref{schlieren} (left) shows the sample schlieren snapshots at different time instants. Fig. \ref{schlieren} (right) shows the processed images in which we can see flow features such as separated shear layer and reattachment with better clarity. Such processed images were used for the analysis of the shock foot unsteadiness which shall be detailed in section \ref{shock_unsteadiness}. 
    
    For unsteady pressure measurements, Kulite XCQ-062 sensors with a rated pressure of 1.7 bar (abs) were employed. Three pressure sensors were simultaneously used per experimental run.  0.5 mm drills from the surface of the plate extending for a depth of about 1 mm was used to expose the sensors to the flow. The data were taken at an acquisition rate of 500 kilosamples/s for 2 s run time using NI-cDAQ model 9185 and NI-9222 module which can accommodate four signal ports at a time. The locations for the Kulite sensors in the current study have been decided based on the zones characterized using the mean pressure measurements in \cite{Sourabh}. Measurements were made along the centreline, the lines located at a distance of 0.87h and 1.5h from the centreline, as shown in fig. \ref{schematic}(c) previously. The positions have been determined such that each distinct pressure zone has a sensor port associated with it. In order to study the shock spectrum and the coherence in shock motion along the span, the pressure ports along the mean shock line were placed at the centreline, as well as at spanwise distances of 0.87h and 1.5h from the centreline. Since the mean shock and the separation bubble are associated with frequency range of few hundreds to about 3-4 kHz (as seen from the literature), all the pressure signals were low-pass filtered at a frequency of 17 kHz before the analyses.

\subsection{DMD of schlieren images and Statistical analyses of pressure}
Dynamic Mode Decomposition (DMD) decomposes the flow-field into spatio-temporal modes that are ordered with respect to the dominant frequency content. It can be thought of as a coupling between singular value decomposition in space and fourier transform in time domain. The current work uses the decomposition procedure described by Kutz et al. \cite{Kutz}. The fluctuation of the pixel intensities in 8000 continuous schlieren snapshots have been used to carry out the DMD analysis. 

The unsteady pressure analysis involves the calculation of properties such as mean and r.m.s of the surface pressure signals along with uncertainty quantification. The uncertainties were calculated by fitting a Student's t-distribution with 95\% confidence limits. Further analyses involved power spectral density estimates, cross correlation and coherence quantification using the MATLAB functions \emph{pwelch}, \emph{crosscorr} and \emph{mscohere} respectively. The PSD's were premultiplied and normalized by the corresponding variance of the signal. A hamming window size of 25000 and 2000 samples with 50\% overlap were used for the PSD calculations of pressure and schlieren shock foot analysis respectively.

\section{Results and discussion}

\begin{figure}[hbt!]
\centering
\includegraphics[width=0.48\textwidth]{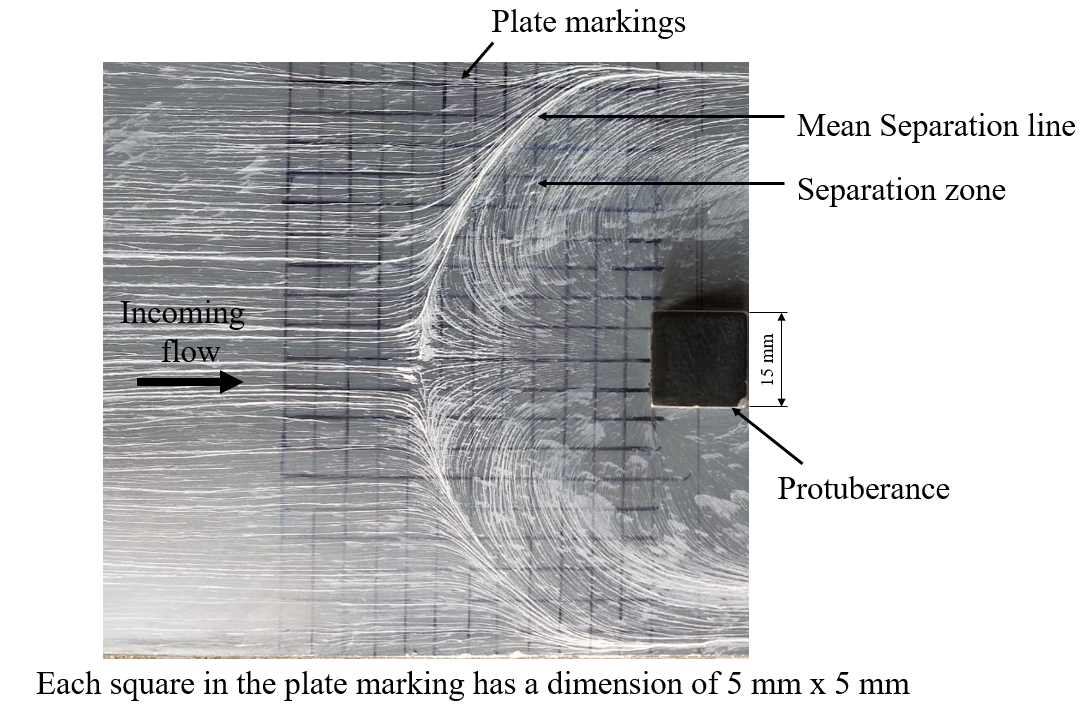}
\caption{Oil flow streakline pattern on the plate surface}
\label{oil_flow}
\end{figure}

The surface oil flow streakline pattern has been shown in fig. \ref{oil_flow} \cite{Sourabh}. It clearly shows the separation line due to the curved shock-boundary layer interaction on the surface of the plate. The plate has been graduated to trace the distance and curvature of the mean separation line with respect to the protuberance. The length of separation along the centreline from the front face of the protuberance was found to be 39 mm ($5.57\delta_x$). We could also clearly observe the streaklines inside the separation region, showing reversal of flow from the protuberance and then curving three-dimensionally outwards, indicating the relieving effect. This effect is not observed in 2-D configurations since the separation bubble remains closed and the flow behaves similarly throughout the span. While we get the mean surface flow features from the oil flow pattern, the length scales of the unsteadiness and the associated frequencies of the shock foot have been quantified using the schlieren images and are discussed in the subsequent section.

\subsection{Shock foot unsteadiness} \label{shock_unsteadiness}

\begin{figure}[hbt!]
\centering
\includegraphics[width=0.48\textwidth]{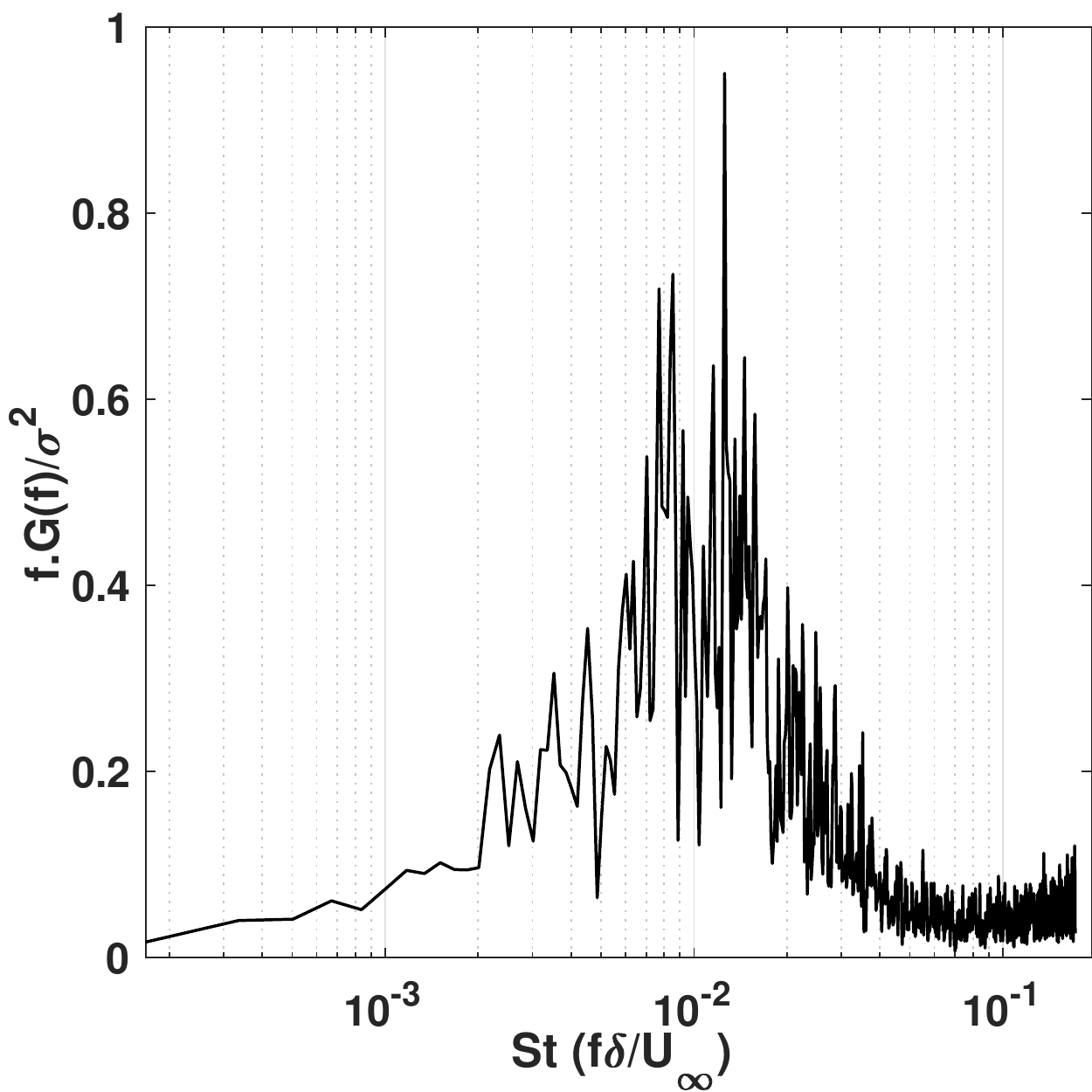}
\caption{Normalized PSD of the shock foot oscillations from schlieren}
\label{scan_line}
\end{figure}

The parameters such as contrast, brightness etc., of the schlieren images were modified in order to better capture the flow features of interest such as the shocks and shear layer structures. To quantify the oscillations of the separation shock, the pixel intensities were scanned along a horizontal line at a distance of $0.9\delta_x$ from the base plate (refer fig. \ref{schlieren}). When scanning from the upstream location, the pixel value corresponding to a sudden intensity drop in each snapshot was noted and the time series formed of these pixel values were used to analyse the unsteadiness. Figure. \ref{scan_line} shows the normalized PSD obtained from the scan line analysis. It is seen that the amplitudes peak around a strouhal number of $10^{-2}$ which agrees with the range widely reported in the literature for 2-D configurations \cite{Clemens}. This analysis could yield the spectrum of shock foot clearly but to capture the structures that contain different characteristic frequencies, dynamic mode decomposition was performed on the processed images whose results are discussed below.

\begin{figure}[hbt!]
\centering
\includegraphics[width=0.48\textwidth]{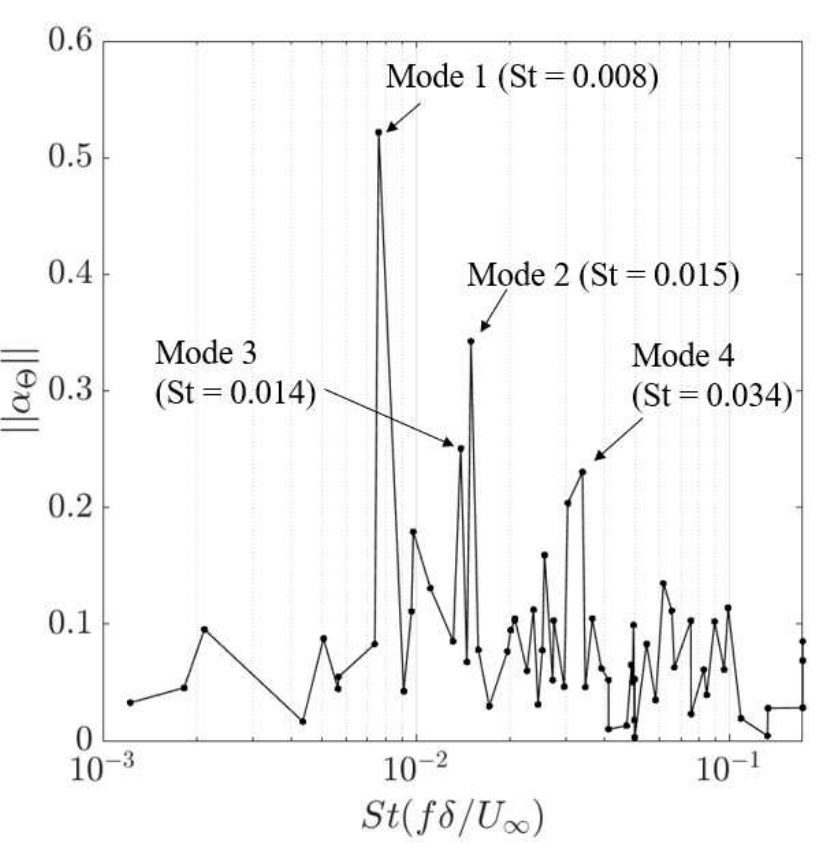}
\caption{DMD spectrum}
\label{spec_ucrp}
\end{figure}

\begin{figure*}
     \centering
     \begin{subfigure}[hbt!]{0.45\textwidth}
         \includegraphics[width=0.98\textwidth]{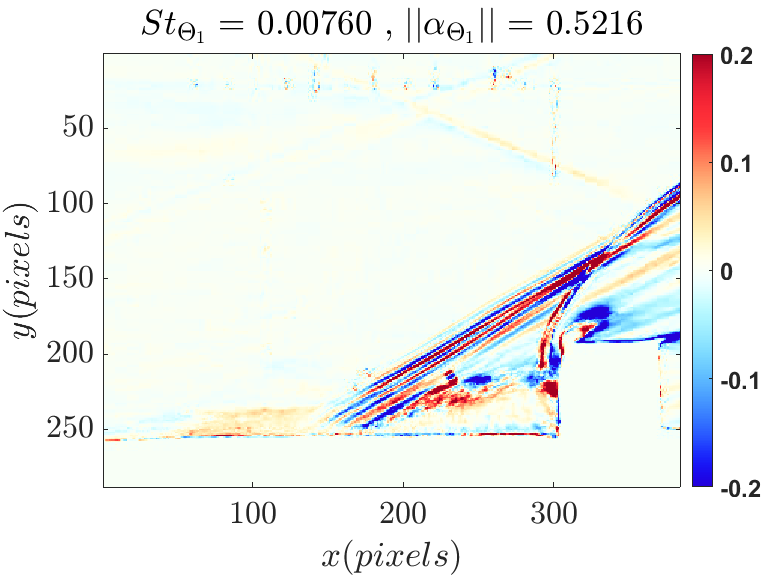}
         \caption{Mode 1}
         \label{mode1_ucrp}
     \end{subfigure}
     \hfill
     \begin{subfigure}[hbt!]{0.45\textwidth}
         \includegraphics[width=0.98\textwidth]{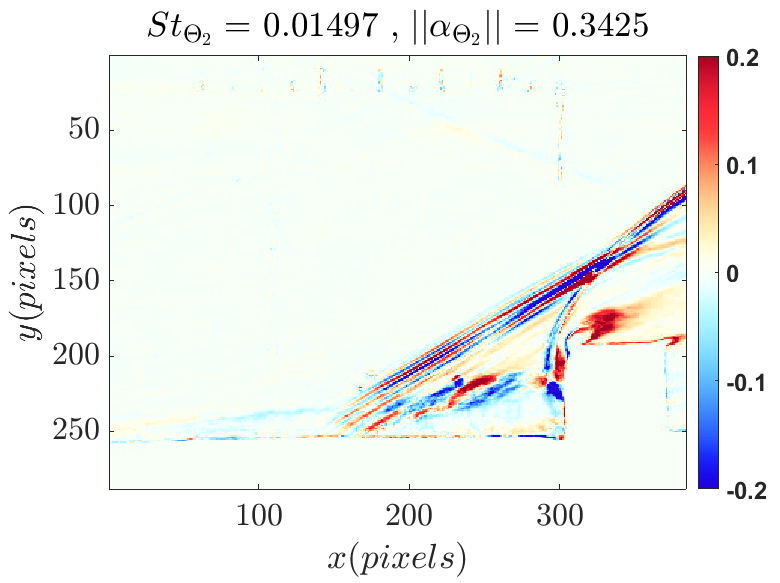}
         \caption{Mode 2}
         \label{mode2_ucrp}
     \end{subfigure}
     \vfill
     \begin{subfigure}[hbt!]{0.45\textwidth}
         \includegraphics[width=0.98\textwidth]{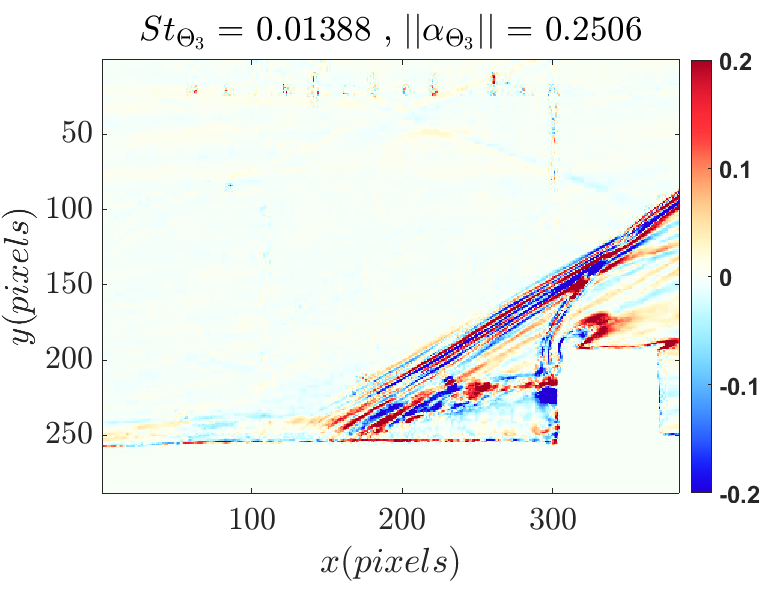}
         \caption{Mode 3}
         \label{mode3_ucrp}
     \end{subfigure}
     \hfill
     \begin{subfigure}[hbt!]{0.45\textwidth}
         \includegraphics[width=0.98\textwidth]{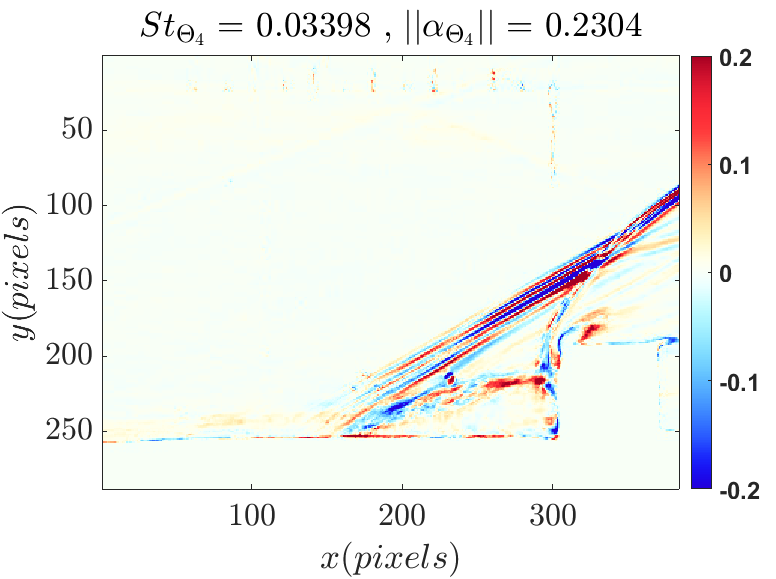}
         \caption{Mode 4}
         \label{mode4_ucrp}
     \end{subfigure}
        \caption{DMD modes}
        \label{modes_ucrp}
\end{figure*}

Figure. \ref{spec_ucrp} shows the plot of DMD spectrum. The amplitudes have been rescaled with bounds 0 and 1. Figure. \ref{modes_ucrp} shows the dominant spatial modes whose frequencies correspond to the first four peak amplitudes (which have also been marked in fig. \ref{spec_ucrp}). It is clearly observed that the separation shock features the most in the dominant mode with a corresponding strouhal number of 0.008, i.e.,  of the order of $10^{-2}$. The dominant mode also highlights some initial shear layer structures close to the separation location. In the higher modes, we could see structures close to the reattachment location and these structures exhibit higher frequencies. For example, in fig. \ref{mode4_ucrp}, the structure close to the reattachment has been captured, which corresponds to a higher Strouhal number of 0.034. Also apparent from the DMD modes is that the separation shock undergoes a to and fro motion as a whole instead of flapping and is in agreement with the observation by Bhardwaj \cite{Bhardwaj} that the separation shock angle remains constant throughout the flow time.

\subsection{Analyses of unsteady pressure}
\label{unsteady_pressure}

\begin{figure}[hbt!]
\centering
\includegraphics[width=0.4\textwidth]{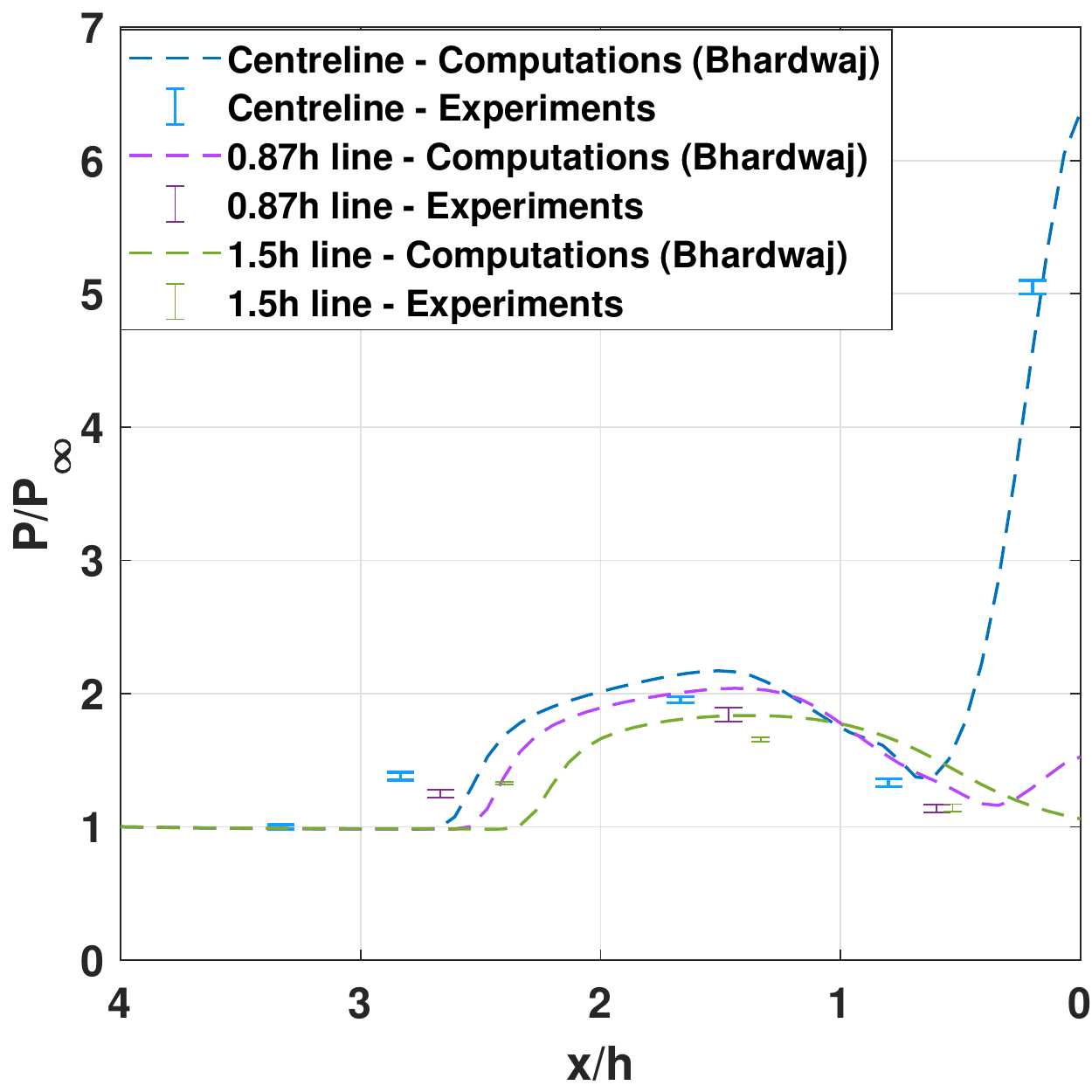}
\caption{Distribution of the normalized mean pressure values along the centreline, 0.87h and 1.5h lines}
\label{mean_stat}
\end{figure}

\begin{figure}[hbt!]
\centering
\includegraphics[width=0.48\textwidth]{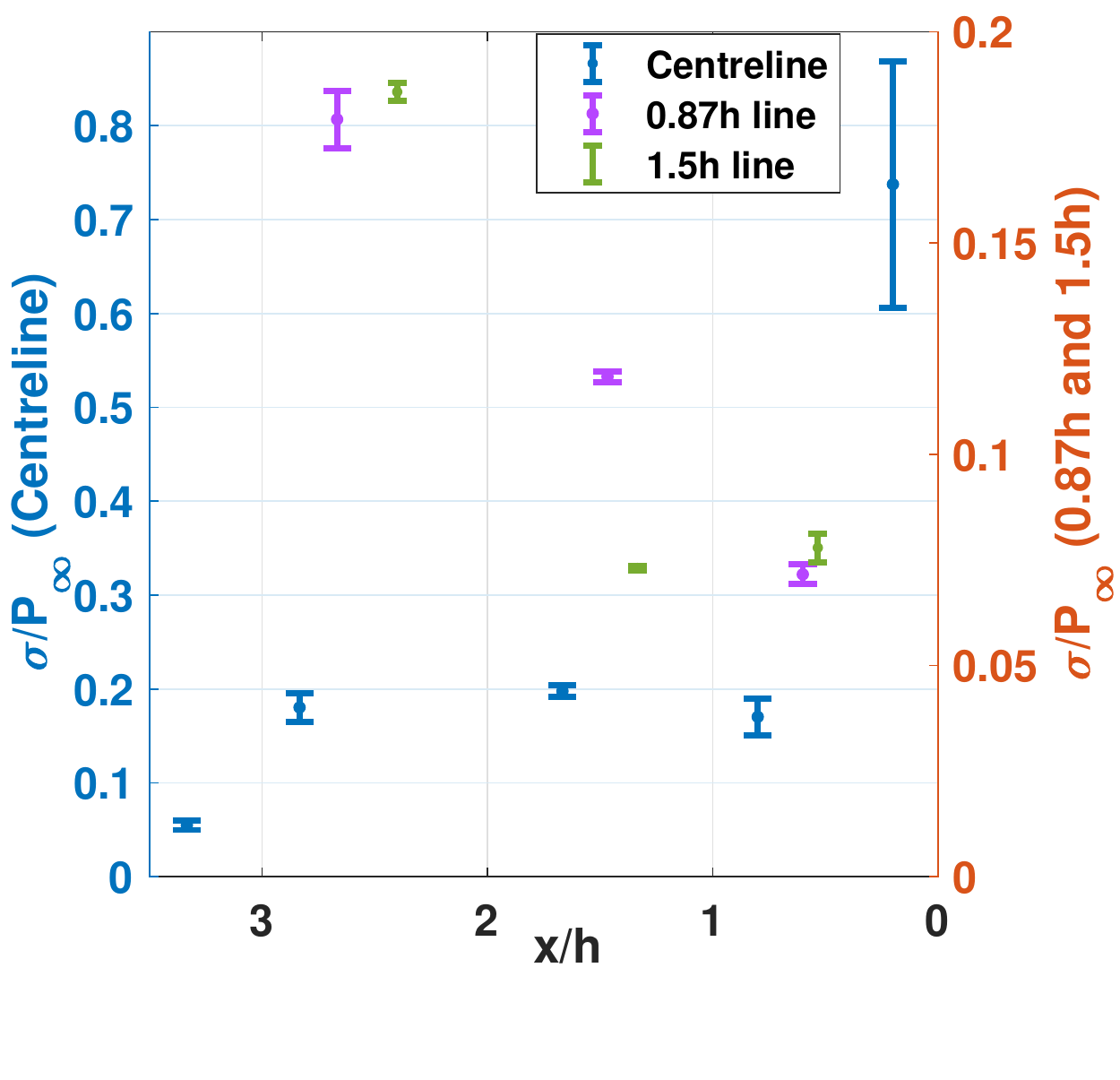}
\caption{Distribution of the normalized r.m.s values of pressure fluctuations along the centreline (left y axis) and 0.87h, 1.5h lines (right y axis)}
\label{std_stat}
\end{figure}

\begin{figure}[hbt!]
\centering
\includegraphics[width=0.48\textwidth]{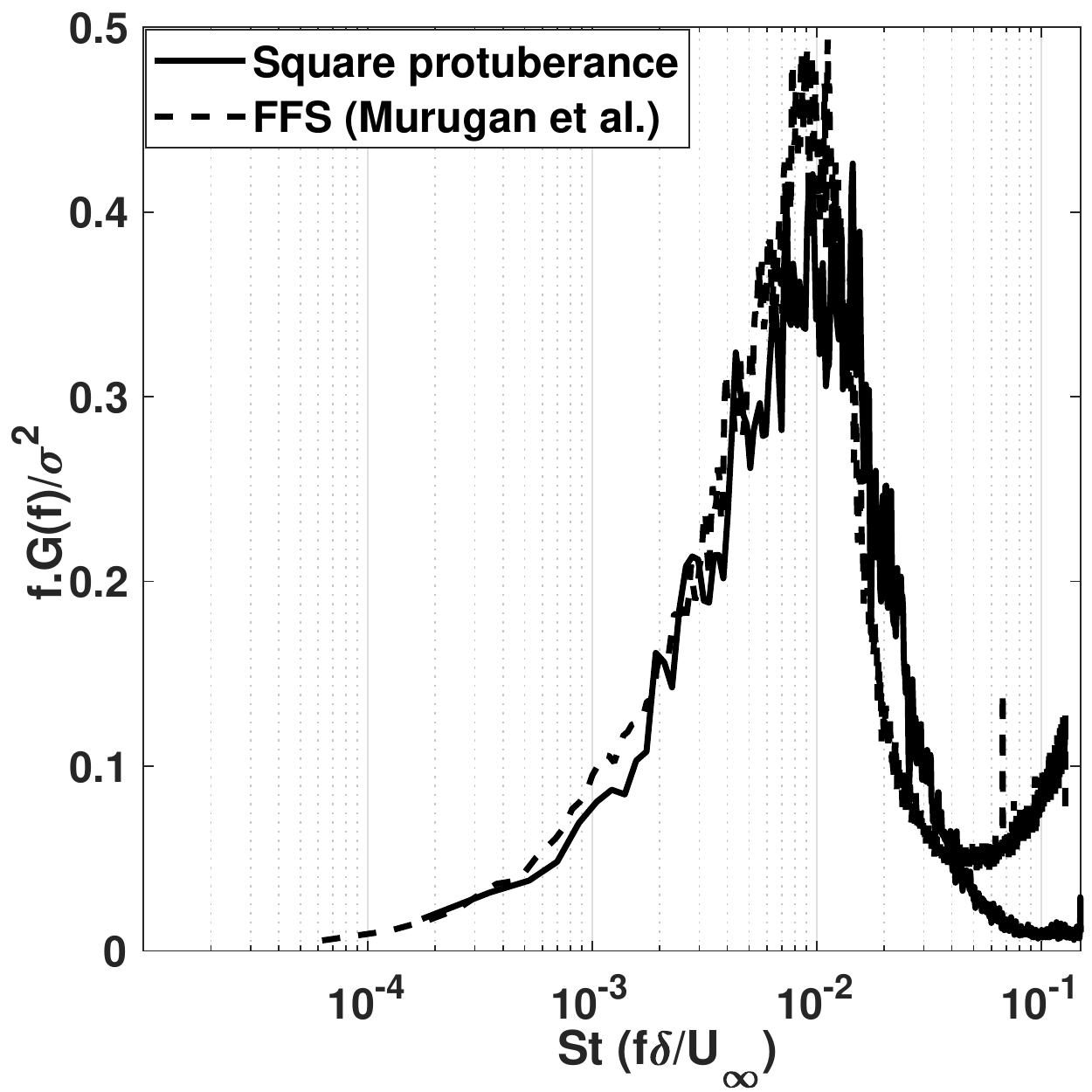}
\caption{Comparison of pressure spectrum at mean shock location between square protuberance and FFS}
\label{ms_comp}
\end{figure}

\begin{figure}[hbt!]
\centering
\includegraphics[width=0.48\textwidth]{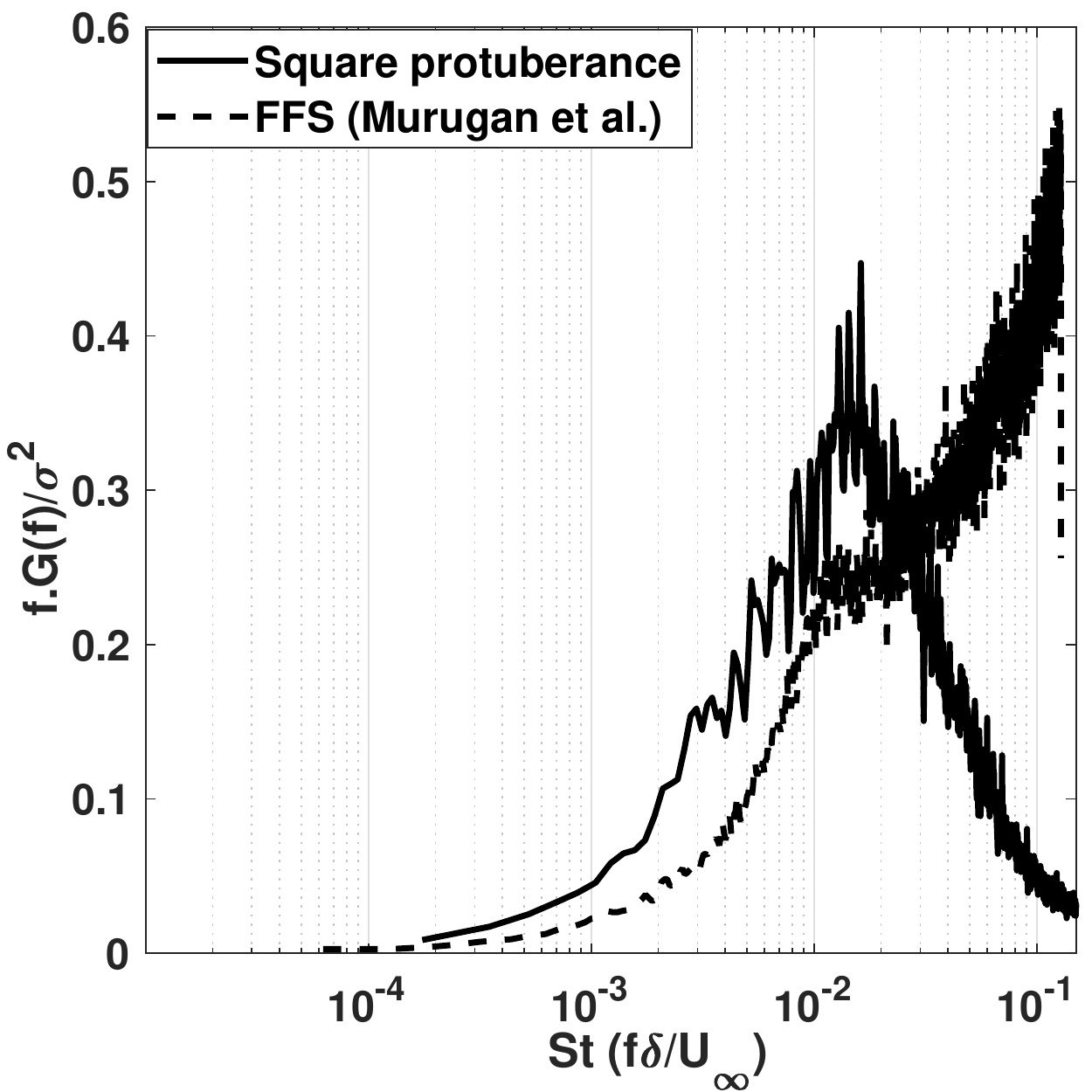}
\caption{Comparison of spectrum at plateau pressure region between square protuberance and FFS}
\label{pp_comp}
\end{figure}

The above analysis using schlieren images quantify the unsteadiness in shock motion only along spanwise centreline. In order to study the flow behaviour along the span and to build the understanding on the overall physics by analysing the nature of different zones of pressure, time-resolved surface pressure measurements were carried out.

Figure. \ref{mean_stat} shows the mean pressure distribution along the centreline, the 13 mm line (0.87h) and the 22.5 mm line (1.5h). Each location of the sensor is associated with several runs ranging from 3 to 10 and the uncertainty has been quantified accordingly. The computational results by Bhardwaj \cite{Bhardwaj} have also been presented alongside the current experimental results in order to provide a reference to the general trend in surface pressure with streamwise distance. It is seen that the results agree quite well, with some deviation in regions of large gradients which may be attributed to the fact that the computations are from steady RANS simulations, which does not account for the shock motion upstream of the mean shock location. The r.m.s values and its associated errors have been presented in fig. \ref{std_stat} for all the three lines of measurements in two different scales (Centreline - left; 0.87h and 1.5h - right). It is observed that the r.m.s values (normalized with respect to free-stream pressure) are less than 0.2 for all the cases except the high pressure zone along the centreline which has a value of 0.74. It is observed that the uncertainty associated with the region is also very high with a value of $\pm0.13$. In general, the uncertainties are lesser along the 13 mm line and 22.5 mm lines compared to that of the centreline.

\subsubsection{Spectral analyses}

\begin{figure*}[!h]
\begin{minipage}[c]{0.32\linewidth}
    \centering
    \includegraphics[width=0.95\textwidth]{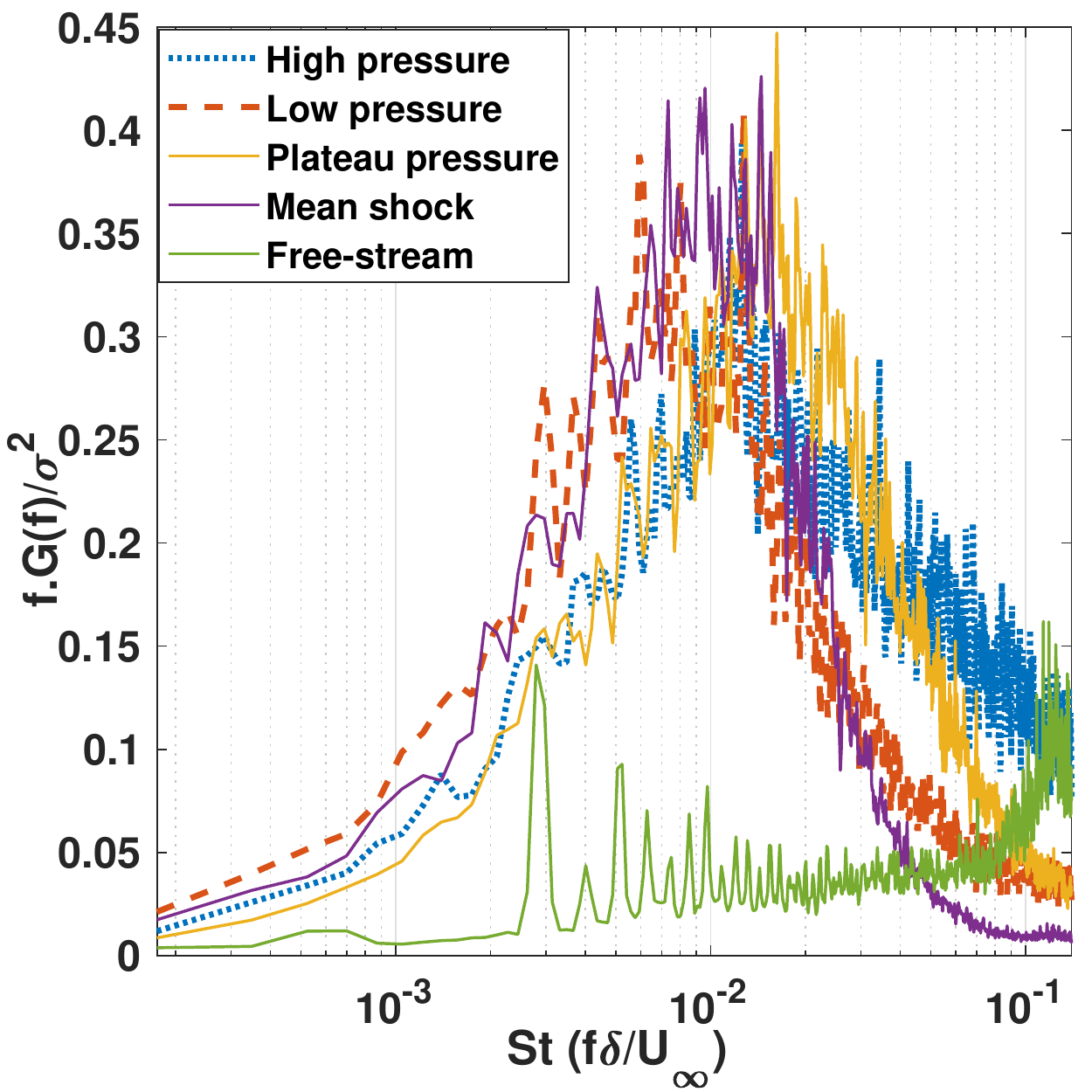}
    \caption{Normalized PSD for different pressure zones along the centreline}
    \label{psd_centreline}
\end{minipage}\hfill
\begin{minipage}[c]{0.32\linewidth}
    \centering
    \includegraphics[width=0.95\textwidth]{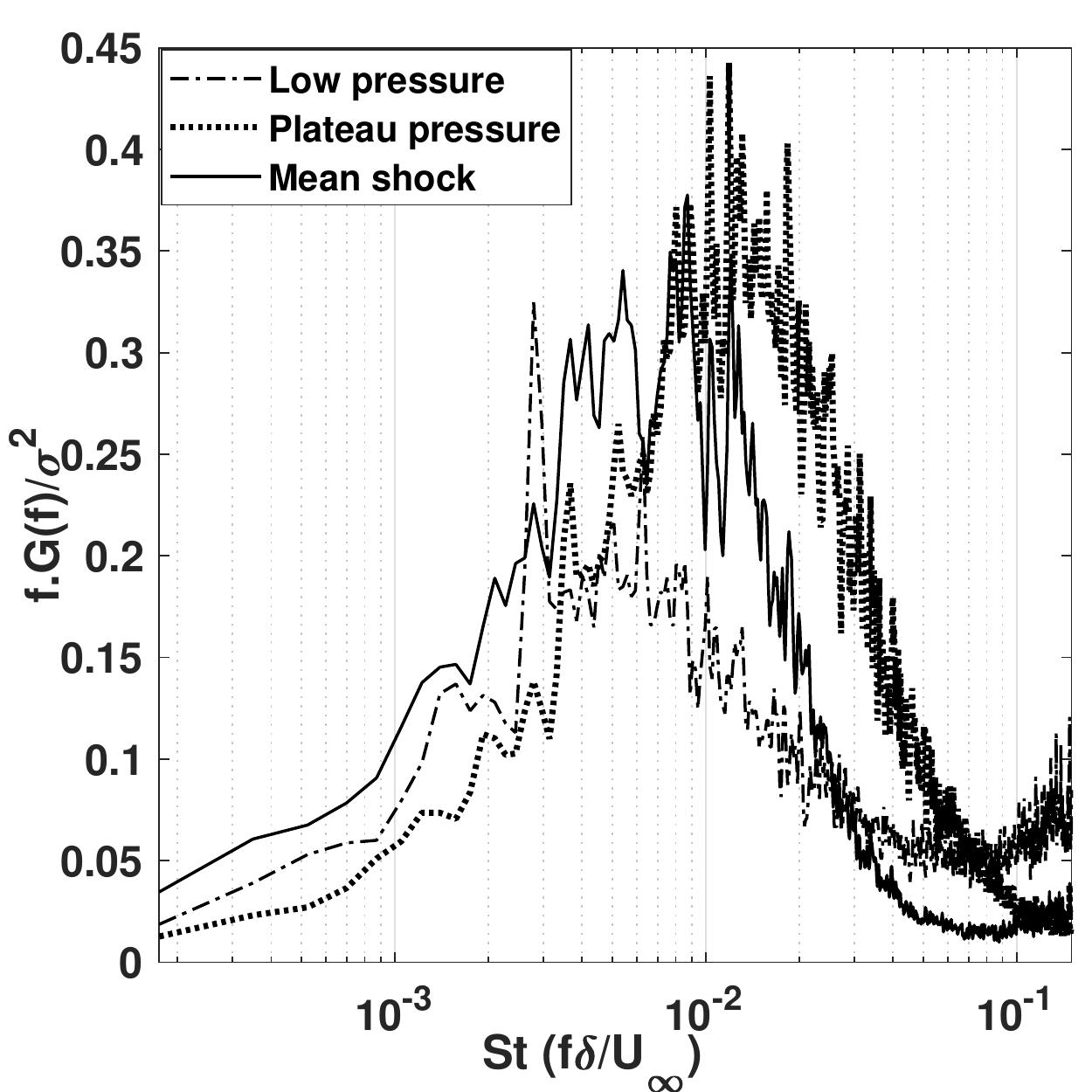}
    \caption{Normalized PSD for different pressure zones along the 0.87h line}
    \label{psd_13mm}
\end{minipage}
\begin{minipage}[c]{0.32\linewidth}
    \centering
    \includegraphics[width=0.95\textwidth]{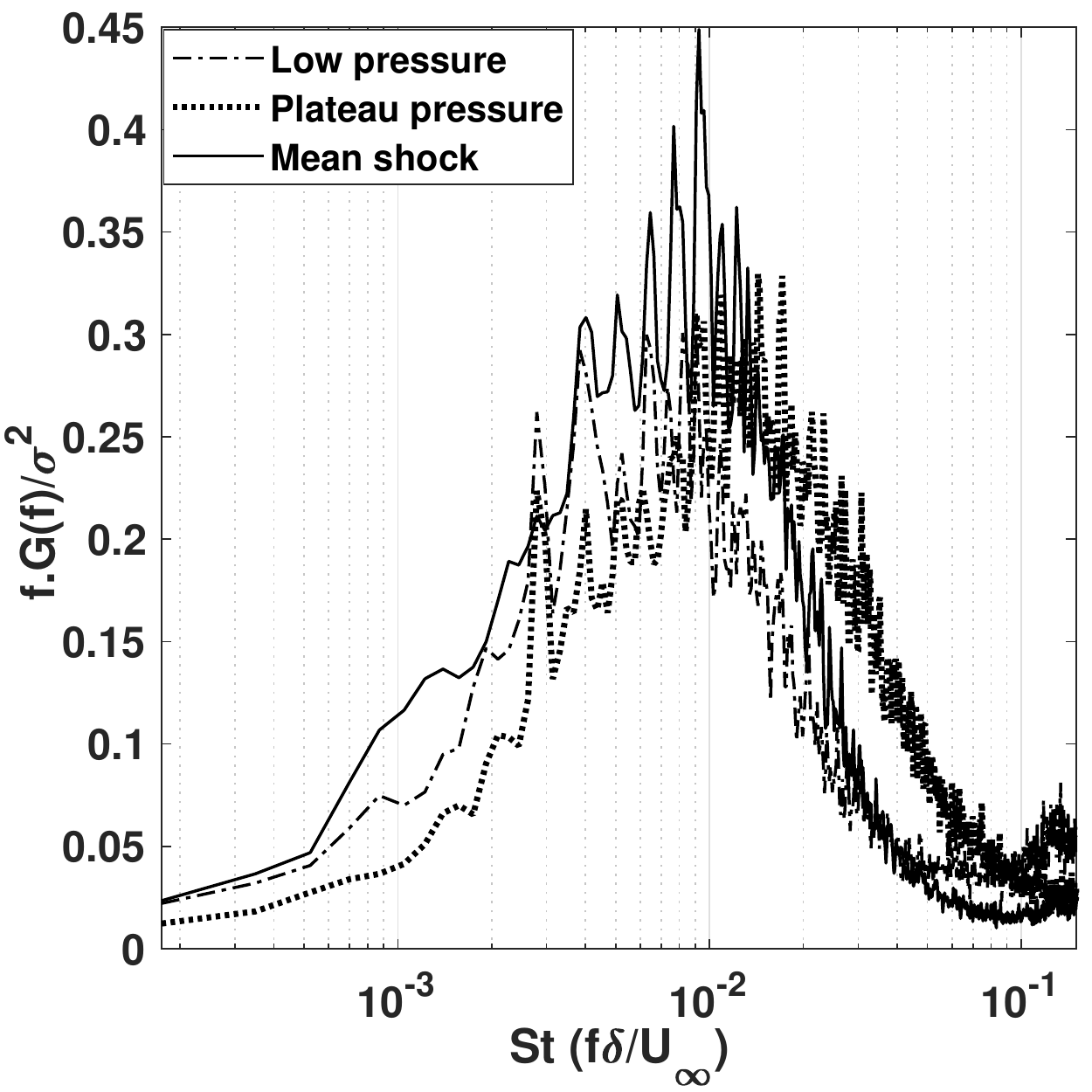}
    \caption{Normalized PSD for different pressure zones along the 1.5h line}
    \label{psd_22.5mm}
\end{minipage}
\end{figure*}

\begin{figure}[hbt!]
\centering
\includegraphics[width=0.48\textwidth]{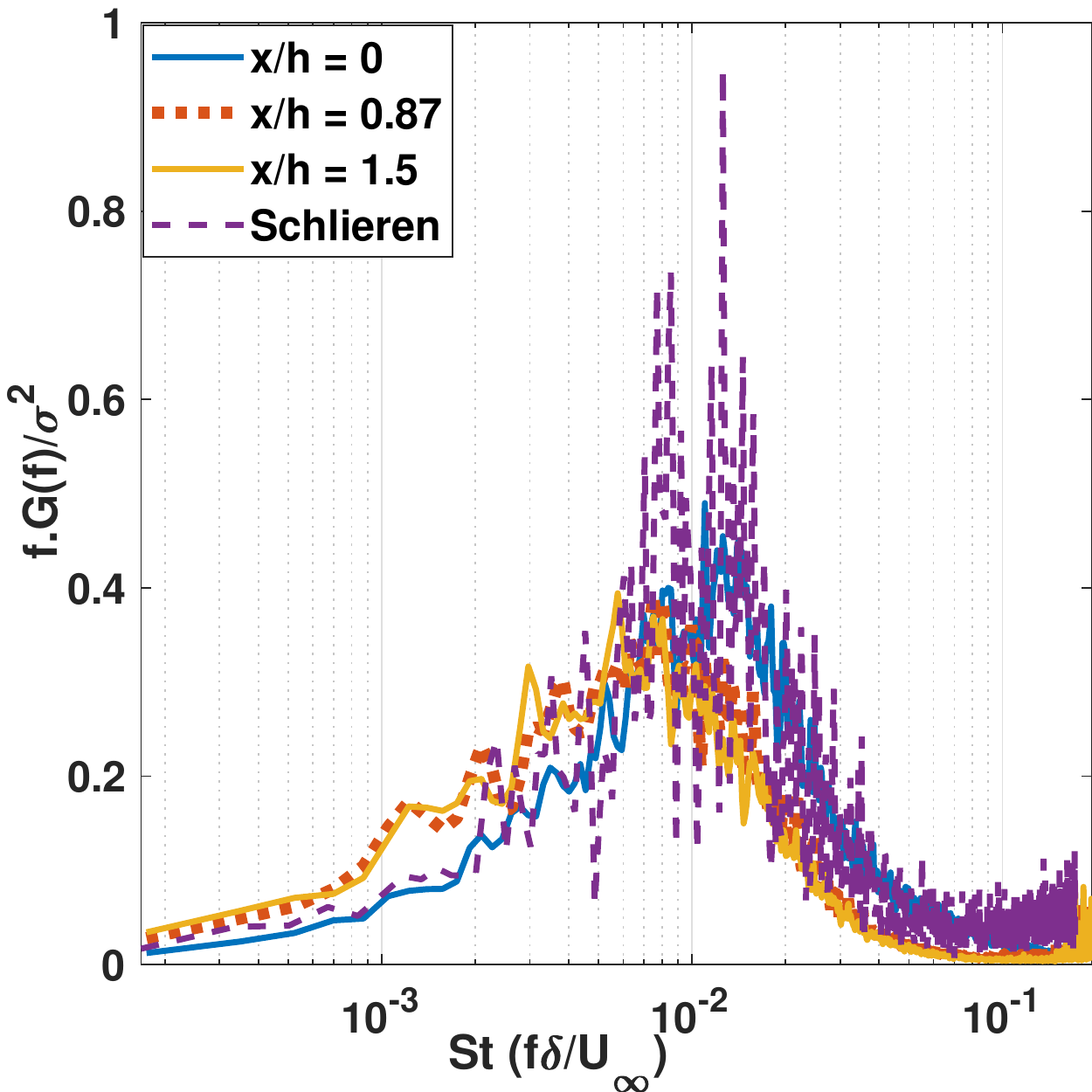}
\caption{Pressure spectra at mean shock foot position for different spanwise locations}
\label{ms_spec}
\end{figure}

The normalized power spectral density of unsteady pressure for the mean shock foot location and the plateau pressure zone obtained for the square protuberance has been compared with the case of Forward Facing step studied by Murugan et al. \cite{Murugan}. Figure. \ref{ms_comp} shows that the pressure spectra at the shock foot location for the 3-D and 2-D (FFS) cases almost coincide with each other with a peak Strouhal number of $10^{-2}$, which represents the low frequency of the separation shock. On the other hand, the spectrum of pressure at the plateau pressure region (fig. \ref{pp_comp}) differs widely. FFS has a rising trend till a Strouhal number of 0.1 with a local plateau around a St of 0.01 whereas the spectrum for the 3-D case peaks around a St of 0.015 ($\sim 1300 Hz$) and then drops. This peak frequency is still higher when compared with that of the shock oscillations.

The power spectral density for the pressure signals along the centreline have been presented in fig. \ref{psd_centreline}. It is clearly seen that the amplitudes of the free-stream signal are considerably low throughout the range since the incoming turbulent eddy scales are associated with much higher frequencies. As discussed earlier, the mean shock location shows a narrow frequency peak around a St of $10^{-2}$ as with the results obtained from the schlieren analysis presented earlier. It is observed that the high pressure region exhibits peak at around the same frequency, though it exhibits considerable amplitudes for higher frequency content too when compared to that of the mean shock location. The spectrum of the low pressure region, which is roughly below the core of the horseshoe vortex, closely follows that of the mean shock. The plateau pressure has a peak at relatively higher frequencies close to 1300 Hz and then starts to drop.

The spectra along 0.87h and 1.5h lines show some interesting results (refer fig. \ref{psd_13mm}). These lines have no high pressure zone since the stagnation due to the obstacle is not strongly felt along the line. The plateau pressure along 0.87h line too shows peaks at relatively higher frequencies (around $St \sim 0.013$) compared to the mean shock and it starts to drop close to $St \sim 0.02$. The low pressure zone shows widely distributed broadband spectrum peaking at a lower frequency between 250 and 700 Hz (which is also significantly lower when compared with the centreline) and then drops gradually. The spectra along the 1.5h line (refer fig. \ref{psd_22.5mm}) also shows a qualitatively similar trend. The plateau pressure spectrum along the 1.5h line has peaks from 800 Hz ($St \sim 0.009$) till 1500 Hz ($St \sim 0.017$), while the low pressure region shows broad peak region between 350 Hz and 800 Hz. 

The mean shock spectrum shows an important trend when we move away from the centreline. The comparison of the mean shock spectrum at different locations is shown in fig. \ref{ms_spec}. It is observed that the schlieren shock foot spectrum matches very well with the centreline pressure spectrum. As we move sideways, the peaks seem to shift towards lower frequency. It is observed that the spectrum of shock foots at 0.87h and 1.5h exhibits peaks around 650 Hz ($St \sim 0.0075$), whereas the centreline shock foot also exhibits similar frequencies but the amplitude remains higher till a strouhal number of 0.015 and then decreases. This possibly indicates that the shock foot along the sides oscillate at a low frequency of around 650 Hz but as we move towards the centre, some additional high frequency content is also seen in the shock foot unsteadiness. 

\begin{figure*}[!h]
\begin{minipage}[c]{0.48\linewidth}
    \centering
    \includegraphics[width=0.95\textwidth]{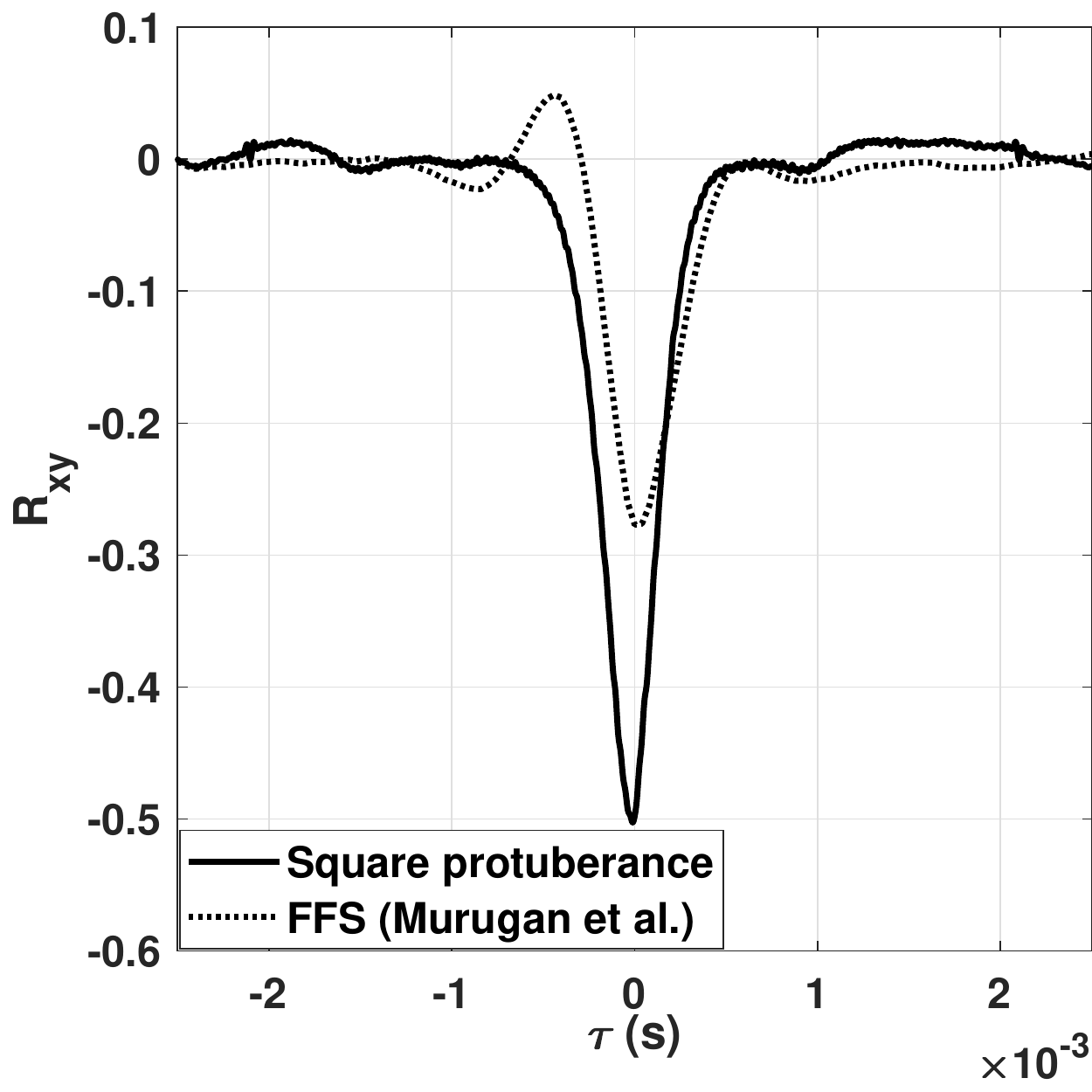}
    \caption{Comparison of cross correlation between mean shock (MS) and plateau pressure regions (PP) for square protuberance and FFS}
    \label{corr_3d_2d_comp}
\end{minipage}\hfill
\begin{minipage}[c]{0.48\linewidth}
    \centering
    \includegraphics[width=0.95\textwidth]{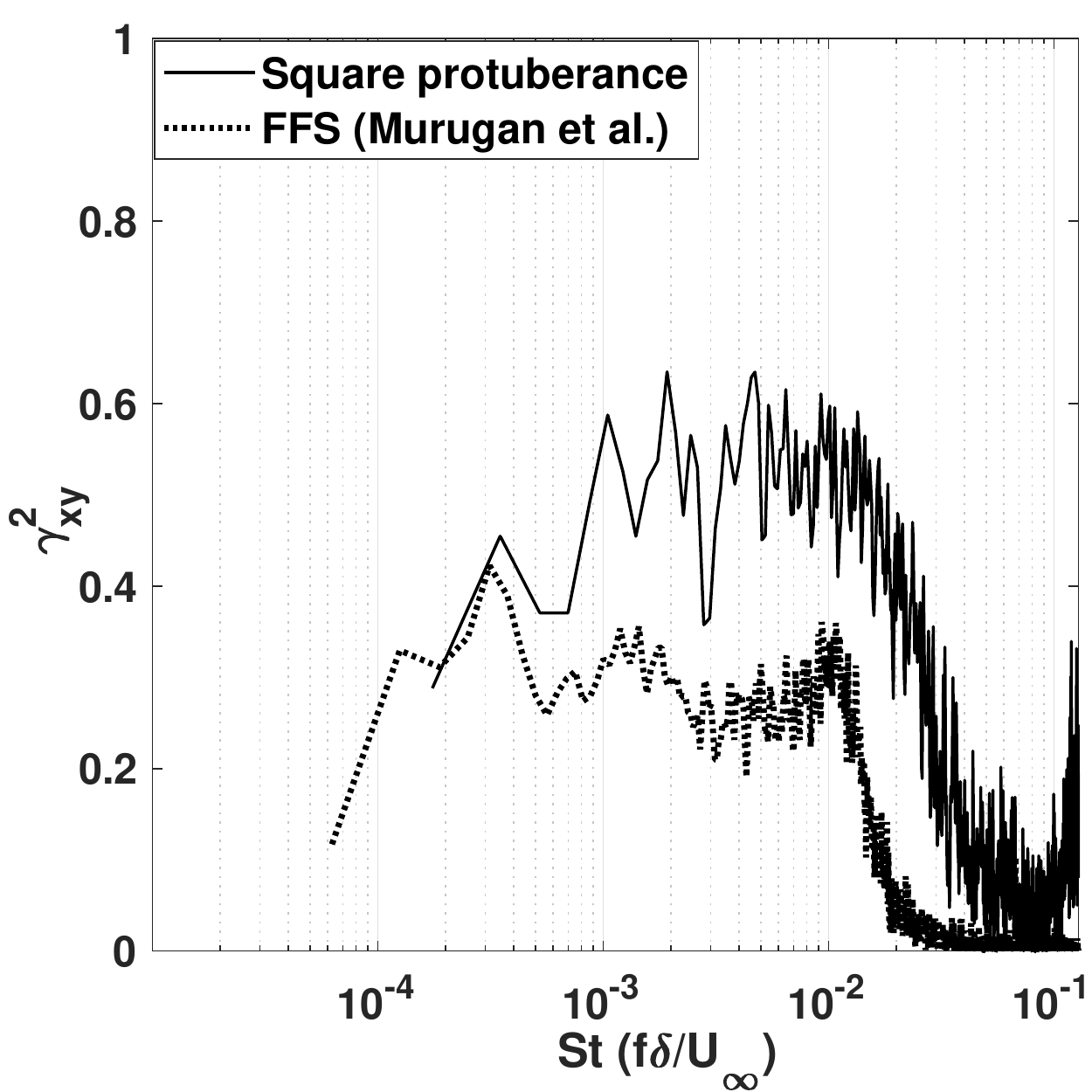}
    \caption{Comparison of magnitude-squared coherence between mean shock (MS) and plateau pressure (PP) regions for square protuberance and FFS}
    \label{coh_3d_2d_comp}
\end{minipage}
\end{figure*}

\subsubsection{Correlation and coherence analyses}

\begin{figure*}[!h]
\begin{minipage}[c]{0.48\linewidth}
    \centering
    \includegraphics[width=0.95\textwidth]{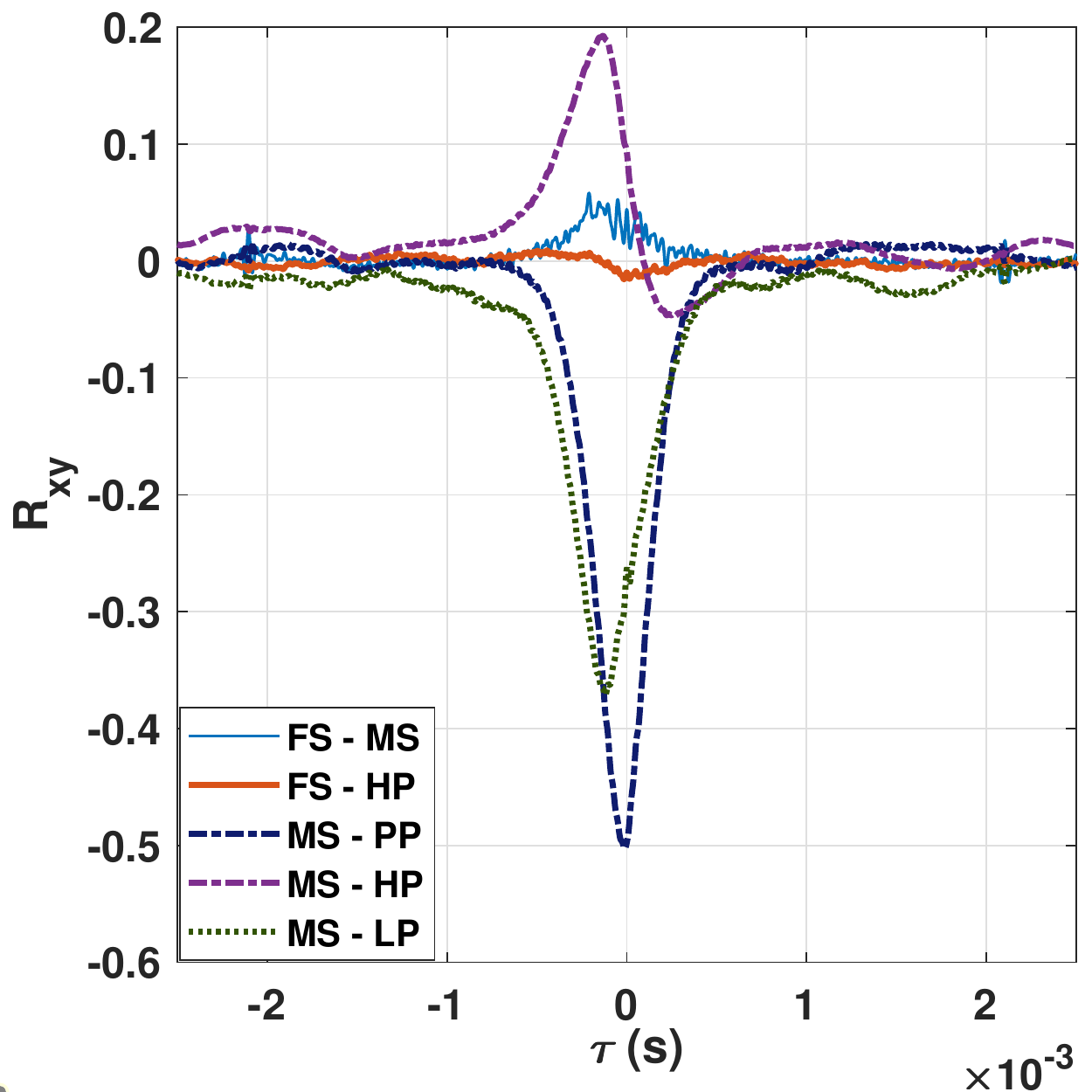}
    \caption{Cross correlation coefficient vs time lag between different zones along the centreline}
    \label{corr_centreline}
\end{minipage}\hfill
\begin{minipage}[c]{0.48\linewidth}
    \centering
    \includegraphics[width=0.95\textwidth]{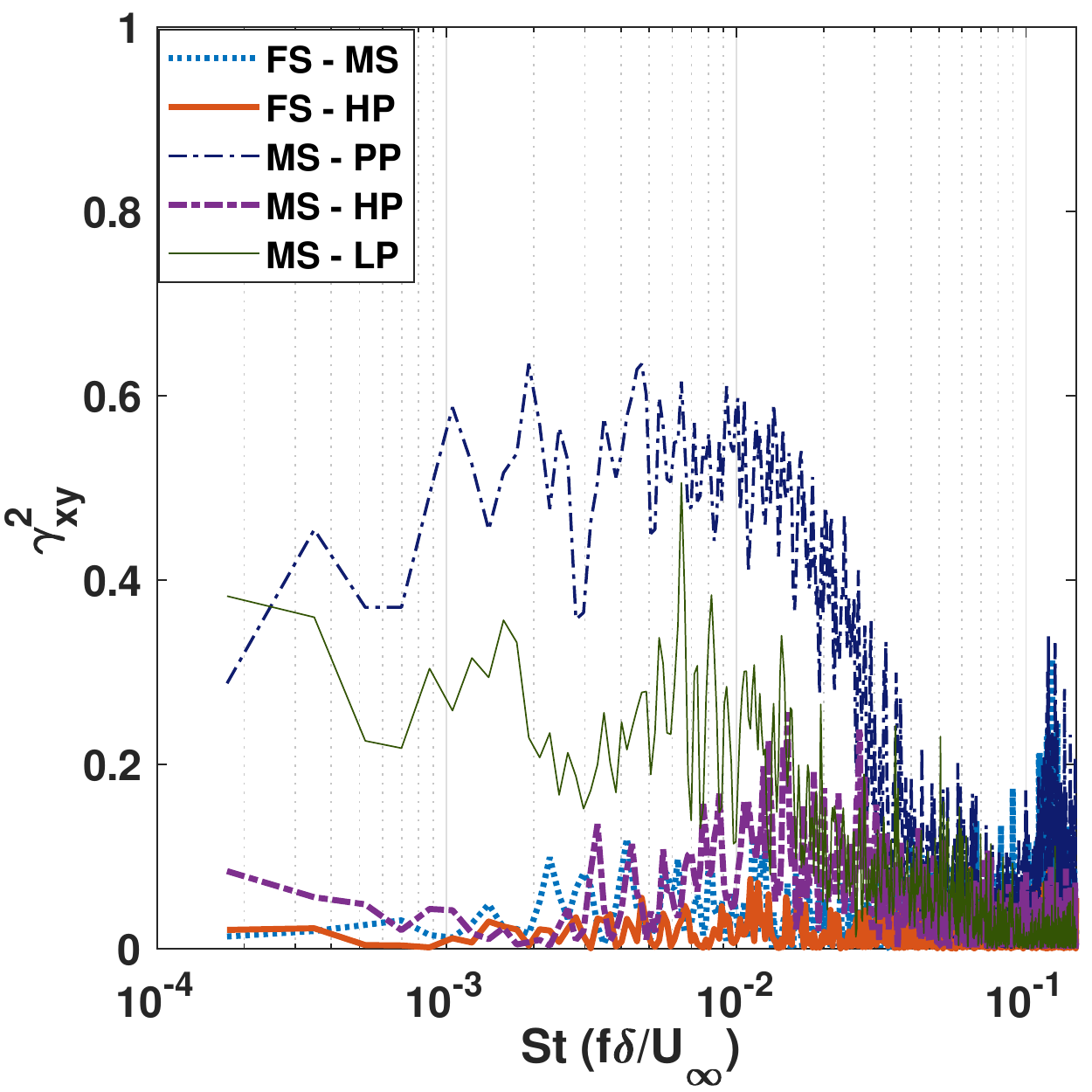}
    \caption{Magnitude-squared coherence vs St between different zones along the centreline}
    \label{coh_centreline}
\end{minipage}
\end{figure*}

The Cross correlation coefficients have been calculated for different time lags between Mean Shock (MS) and Plateau Pressure zones (PP) and are compared with the 2-D protuberance \cite{Murugan}. The comparison is plotted in fig. \ref{corr_3d_2d_comp} with a time lag window ranging from -2.5 ms to 2.5 ms. All the plots have been labelled such that a peak at a positive time lag between regions `1-2' would indicate that the event at 2 follows that of 1 (Here, 'MS' is 1 and 'PP' is 2). Correlation between the mean shock and the plateau pressure region for the square protuberance shows a huge negative peak of -0.5 at a very small time lag of about $-10 \mu s$. This depicts the strong dependence of the shock foot oscillations with the separated region in the immediately vicinity. This has been reported earlier in many works \cite{Piponniau,Murugan} for 2-D configurations. The negative peak implies that a pressure rise in the mean shock location is accompanied by a decrease in pressure in the plateau region and vice-versa. This is because, when the shock foot moves upstream resulting in pressure rise in the transducer, the bubble expands to a larger size thus decreasing the pressure in the plateau region.  We could observe that the FFS case shows a strong negative correlation but with value much less than that of the 3-D counterpart. The time lag is seen to be almost close to $0 \mu s$ in the 2-D case whereas it is $-10 \mu s$ for the square protuberance, whose difference arises from the resolution of the data presented, which in turn depends on the data acquisition rate (Data for the present case has been acquired at 500 kHz whereas FFS data had been acquired at 25 kHz \cite{Murugan}). The negative time lag indicates that the disturbance is travelling upstream from the plateau region to the shock foot. 

The values of Magnitude-squared coherence vs Strouhal number between mean shock and plateau pressure region (refer fig. \ref{coh_3d_2d_comp}) for the square protuberance show a broad peak from 200 - 1000 Hz and then starts to drop in the close vicinity to St of $10^{-2}$. The plot also qualitatively follows that of the FFS case, however with a relatively higher peak value of around 0.6.
 
Cross correlation coefficients were calculated between the following regions along the centreline: Free-stream (FS), Mean shock (MS), Plateau pressure (PP), Low pressure (LP) and High pressure (HP) (refer fig. \ref{corr_centreline}). The correlation coefficients between the free-stream region and the mean shock as well as the high pressure zone are very low with values less than 0.06. This is because of the fact that the incoming turbulence plays a minimal role in strong interactions. A negative peak of -0.37 was observed between the mean shock and the low pressure region with a time lag of about $-120 \mu s$, meaning that the region exhibits qualitatively similar correlation when compared to that with the plateau pressure region.

A positive peak of 0.2 at a negative time lag of $-130 \mu s$ was observed between the mean shock and the high pressure zone close to the protuberance indicating that a pressure drop in the high pressure zone is associated with a downstream shock sweep and vice-versa. It is observed that the shock-plateau, shock-low pressure and the shock-high pressure region correlations peak at negative time lag implying that the event at the former location follows the latter. These data suggest that the disturbances might be travelling upstream inside the bubble. The absolute value of the time lag is observed to increase with increase in distance between the pressure zones.

\begin{figure*}[!h]
\begin{minipage}[c]{0.48\linewidth}
    \centering
    \includegraphics[width=0.95\textwidth]{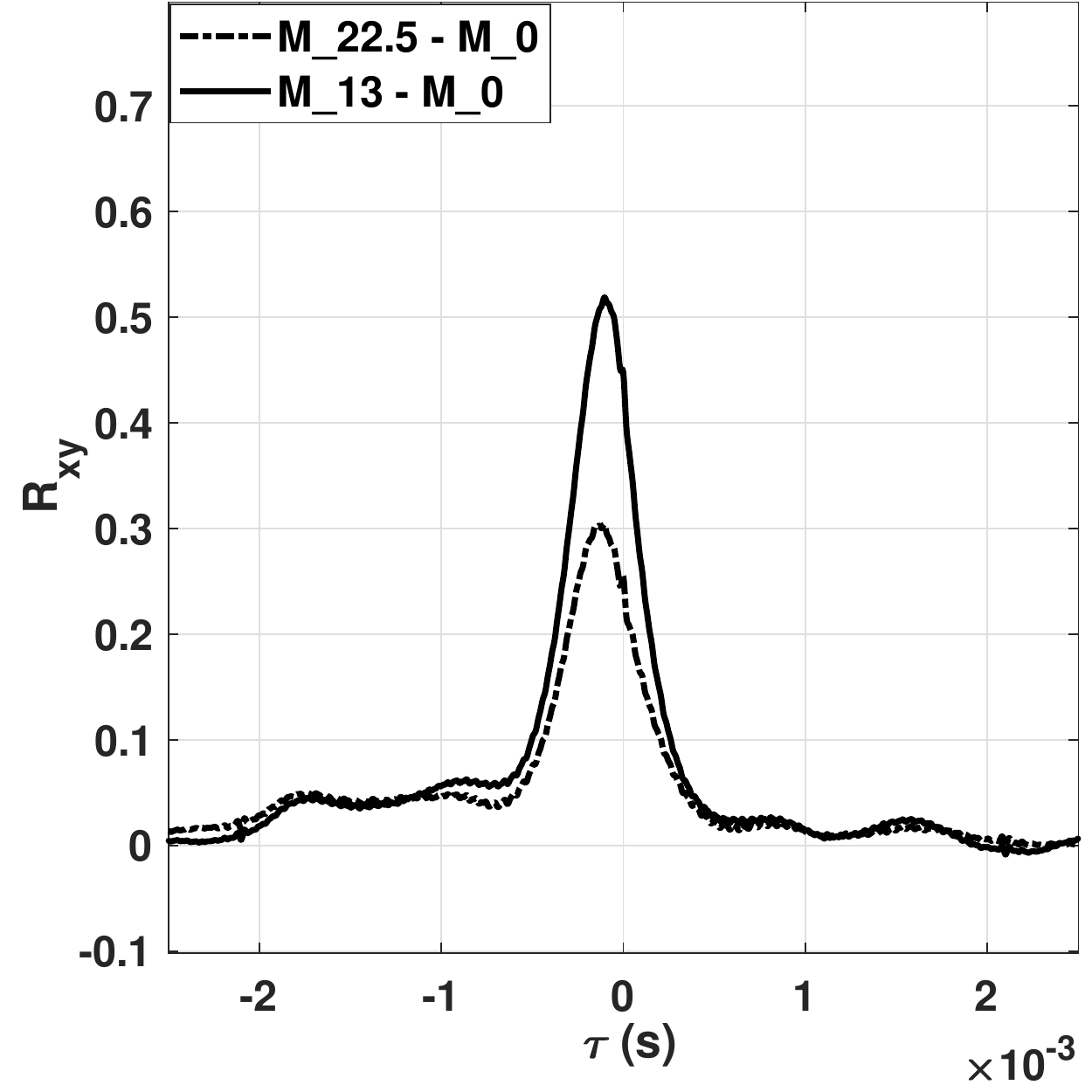}
    \caption{Correlation between different spanwise mean shock foot locations}
    \label{cor_ms}
\end{minipage}\hfill
\begin{minipage}[c]{0.48\linewidth}
    \centering
    \includegraphics[width=0.95\textwidth]{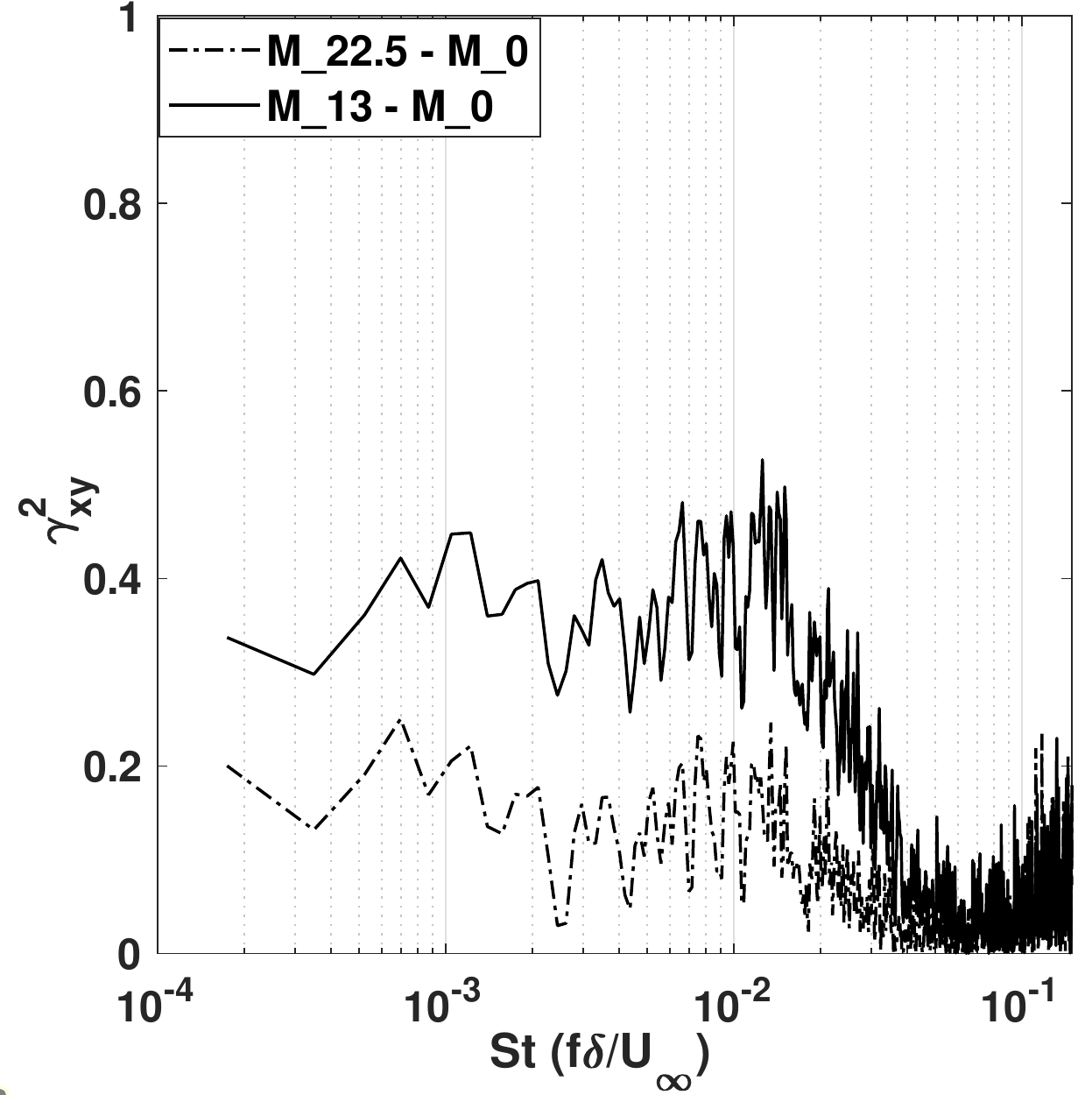}
    \caption{Coherence between different spanwise mean shock foot locations}
    \label{coh_ms}
\end{minipage}
\end{figure*}

The magnitude-squared coherence has been plotted between the different regions along the centreline in fig. \ref{coh_centreline}. As anticipated, the free-stream region is not related to the mean and high pressure signals and shows much lesser values. It is observed that the mean shock is related to the high pressure region in the higher frequency side ($>$ 1 kHz). The mean-low pressure region shows peak values around St of 0.006 (500 Hz) and is narrow compared to the mean-plateau coherence. These results also qualitatively agree with the individual power spectral densities of the corresponding regions.

Further analyses involved placing the sensors at mean shock locations at different spanwise points. Figure. \ref{cor_ms} shows the correlation between the mean shock positions at 0 mm, 13 mm and 22.5 mm from the centreline. The correlation values are significant and is observed that the correlation of the centreline mean shock ($M\_{0}$) with that of the mean shock foot at 22.5 mm line ($M\_{22.5}$) is less than that with the mean shock foot at 13 mm line ($M\_{13}$), since the former location is farther compared to the latter. As anticipated, all values were positive meaning that at on average, shock moves together to and fro along the span. It is evident from the negative time lag values that the shock foot at the centreline, always precedes the sides. It is also observed that the shock foot at the 13 mm line has a lesser time lag ($-104 \mu s$) when compared to that of the 22.5 mm line ($-134 \mu s$) meaning that the shock movement occurs in the order of centreline to sides (i.e., the centreline shock foot moves first, followed by 13 mm and so on). The coherence plots (refer fig. \ref{coh_ms}) follow a similar trend with 13 mm line showing higher coherence (with the centreline shock foot) when compared to that of the 22.5 mm line and contains broad frequency peaks till about 1300 Hz ($St \sim 0.015$).

\section{Conclusions}
The unsteadiness associated with the 3-dimensional shock induced separation due to square faced protuberance has been studied using time-resolved schlieren and unsteady pressure measurements. Dynamic mode decomposition of schlieren snapshots identifies the low frequency to and fro oscillation of the separation shock foot (with $St \sim 0.01$) at the centreline as the most dominant mode which agrees well with the corresponding shock foot spectrum from schlieren and pressure measurements. 

The spectrum of pressure at centreline mean shock location matched well with the 2-D counterpart whereas the plateau pressure spectrum differed widely. The pressure spectrum at mean shock, plateau pressure and low pressure regions differ as we move along the span from the centreline. Towards the sides, the range of peaks in the plateau pressure as well as low pressure zones shifts to lower frequencies, with the spectrum becoming relatively more broadband for the low pressure zone compared to the corresponding spectrum at centreline. It is seen that the whole shock moves with a common low frequency of around 650 Hz though the centreline exhibits some additional high frequencies too in the range of 900 Hz. The cross correlations clearly identify the independence of the shock and the bubble unsteadiness with the fluctuations in the free-stream. It is also observed that the shock foot and the plateau pressure region are strongly correlated. It was clearly identified that the centreline shock foot precedes that at the sides and on average, the whole shock foot moves together.

\section*{Acknowledgments}
The authors are grateful to Dr. G. Rajesh and his students for helping with some of the experimental facilities. We would also like to thank Mr. Abhishek Kumar, Mr. Hemanth Chandravamsi, Mr. Vayala Siva and Mr. Sivaprasad for the healthy discussions and their constant support in experimentation. The research work is supported by the Science and Engineering Research Board of the Department of Science and Technology, Government of India, SERB grant no.
SRG/2019/001793.

\bibliography{sn-article}


\end{document}